\documentclass[12pt]{article}
\newcommand{\R}{{\rm R}}
\newcommand{\C}{{\rm C}}

\newcommand{\SO}{{\rm SO}}
\newcommand{\SL}{{\rm SL}}
\newcommand{\Sp}{{\rm Sp}}
\newcommand{\ISO}{{\rm ISO}}
\newcommand{\SU}{{\rm SU}}
\newcommand{\U}{{\rm U}}
\newcommand{\E}{{\rm E}}
\newcommand{\Hol}{{\mathcal H}}
\newcommand{\Spin}{{\rm Spin}}

\newcommand{\beq}{\begin{equation}}
\newcommand{\eeq}[1]{\label{#1}\end{equation}}
\newcommand{\bea}{\begin{eqnarray}}
\newcommand{\eea}[1]{\label{#1}\end{eqnarray}}

\newcommand{\fft}[2]{{\frac{#1}{#2}}}
\newcommand{\ft}[2]{{\textstyle{\frac{#1}{#2}}}}
\ifx\Bbb\undefined
\else
\renewcommand{\R}{{\mathbb R}}
\renewcommand{\C}{{\mathbb C}}
\fi
\ifx\ltimes\undefined
\newcommand{\ltimes}{{\kern3pt\hbox{\vrule width 0.4pt height 5.30pt depth
.0pt}\kern-1.76pt\times\kern1pt}}
\fi
\begin{document}

\begin{titlepage}
\begin{flushright}
MCTP-04-18\\
hep-th/0403160
\end{flushright}

\vspace{15pt}

\begin{center}
{\large\bf Erice lectures on ``The status of local 
supersymmetry''\footnote{Lectures given at the International School of 
Subnuclear Physics, Erice, August 2003. Research supported in part by 
DOE Grant DE-FG02-95ER40899.}}

\vspace{15pt}

{M.~J.~Duff\footnote{mduff@umich.edu}}

\vspace{7pt}

{\it Michigan Center for Theoretical Physics\\
Randall Laboratory, Department of Physics, University of Michigan\\
Ann Arbor, MI 48109--1120, USA}

\end{center}

\begin{abstract}

In the first lecture we review the current status of 
local supersymmetry.  In the second lecture we focus on D=11 
supergravity as the low-energy limit of M-theory and pose the questions: 
(1) What are the D=11 symmetries? (2) How many supersymmetries can 
M-theory vacua preserve?

\end{abstract}

\end{titlepage}


\newpage
\tableofcontents
\newpage

\section{Local supersymmetry in supergravity, superstrings and M-theory}
\label{one}

{\it Gravity exists, so if there is any truth to supersymmetry then any
realistic supersymmetry theory must eventually be enlarged to a
supersymmetric theory of matter and gravitation, known as
supergravity.  Supersymmetry without supergravity is not an option,
though it may be a good approximation at energies below the Planck
Scale.}

Steven Weinberg, The Quantum Theory of Fields, Volume III, Supersymmetry

\subsection{Supergravity}
\label{introduction}
\indent

 The organizers of the school requested that I review the status of
``local supersymmetry''.  Since local supersymmetry represents a large
chunk of the last 25 years of research in theoretical high energy
physics, I will necessarily be selective.  Local supersymmetry appears
in supergravity, superstrings, supermembranes and M-theory.  A complete
treatment of strings, branes and M-theory is beyond the scope of these
lectures and they will deal mostly with supergravity.  In my opinion
there are currently four reasons why supergravity is interesting:

1) Ten dimensional and eleven dimensional supergravity respectively
describe the low energy limits of string theory and
M-theory, which represent our best hope for a unification of all
fundamental phenomena: particle physics, black holes and cosmology.
Supergravities in lower dimensions are also important for discussing
compactifications.  Pending such a final theory there are less
sweeping but more tractable uses of supergravity such as:

2) The gauge-theory/supergravity correspondence allows us to use our
knowledge of weakly coupled five-dimensional supergravity to probe
strongly coupled four-dimensional gauge theories such as QCD.

3) Cosmological solutions of supergravity hold promise of explaining
inflation and the current acceleration of the universe.

4) There is still no direct experimental evidence for supersymmetry
but it might be the panacea for curing the ills of non-supersymmetric 
theories of particles and cosmology:

{\it The gauge hierarchy problem}

{\it Electroweak symmetry breaking}

{\it Gauge coupling unification}

{\it Cold dark matter}

{\it Baryon asymmetry}

Let us recall that global supersymmetry unifies bosons and fermions by
requiring that our equations be invariant under a transformation
involving a constant fermionic parameter $\epsilon$ which converts
boson fields $B$ to fermion fields $F$ and vice versa.  Symbolically
\begin{equation}
\delta F= \partial B\epsilon~~~~~~~~~\delta B=\bar{\epsilon}F
\label{epsilon}
\end{equation}
Here $B$ is commuting while $F$ and $\epsilon$ are anticommuting.
There can be up to 4 such supersymmetries in four spacetime dimensions: simple
$N=1$ and extended $N=2$,4.  The maximum spin allowed is $s=1$.  The maximum
spacetime dimension allowed is $D=10$ corresponding to 16 spinor components.

Local supersymmetry means that we allow $\epsilon$ to be a function of the
spacetime coordinates. The massless gauge field associated with local supersymmetry
is a spin 3/2 fermion, the gravitino.  Interestingly enough, local supersymmetry
necessarily implies invariance under general coordinate
transformations and so, as its name implies, the gravitino is the
superpartner of the graviton.  There can be up to 8 such
supersymmetries in four spacetime dimensions: simple $N=1$ and
extended $N=2$,3,4,5,6,8.  The maximum spin allowed is $s=2$.  The
maximum spacetime dimension allowed is $D=11$ corresponding to 32
spinor components.

The status of local supersymmetry is largely the status of
supergravity: the supersymmetric version of general relativity
discovered in 1976.  This is the original reason for the popularity
of supergravity: it provides a natural framework in which to unify
gravity with the strong, weak and electromagnetic forces.  This is the
top-down approach.

Local supersymmetry played a major part in many subsequent developments such
as matter coupling to supergravities, the super Higgs mechanism,
anti de Sitter supergravities, BPS black holes and supersymmetric
sigma-models.  Many of these contributed to the phenomenological
application of supergravity-induced supersymmetry breaking in the
physics beyond the standard model, as well as to the connection
between Yang-Mills theories and supergravity via the AdS/CFT
correspondence.

It is important not only as supersymmetric extension of gravity but has
also had a significant impact on other fields.  In standard general
relativity it has given rise to positive energy theorems and to new
results in the study of black holes, extra spacetime dimensions and
cosmology.

Since local supersymmetry places an upper limit on the dimension of
spacetime, it naturally suggests that we incorporate the
Kaluza-Klein idea that our universe may have hidden dimensions in
addition to the familiar three space and one time.

Since my job is to evaluate the status of local
supersymmetry, I shall not spend much time with introductions.
Rather I wish in this first lecture to explain where it stands in the
grand scheme of things and to what extent the top-down approaches
enumerated in (1)-(3) above and bottom-up approaches of (4) are
compatible.  In this connection, we note that he criterion of chirality
in four dimensions means that only simple $N=1$ supersymmetry could be   
directly relevant to observed particles.  However, such models can
emerge from both simple and extended theories in higher dimensions.

Early discussions of local supersymmetry may be found in the papers 
of Volkov and Soroka \cite{Volkov1,Volkov2}.
Supergravity was introduced by Ferrara, Freedman and van Nieuwenhuizen 
\cite{Ferrara} and by Deser and Zumino \cite {Deser}.  Introductions to 
supersymmetry and supergravity may be found in the books
by Bagger and Wess \cite{Wess}, Gates, Grisaru, Rocek and Siegel
\cite{Gates}, Srivastava \cite{Srivastava}, West 
\cite{West}, Freund \cite{Freund}, Bailin and Love \cite{Bailin} and Weinberg 
\cite{Weinberg}.  See also the Physics Reports of Sohnius 
\cite{soh}, van Nieuwenhuizen \cite{vanN} and Fayet and Ferrara 
\cite{Fayet} and the review by Lykken \cite{Lykken}.

For phenomenological applications of local supersymmetry see the 
lecture of Ellis \cite{Ellis} and the
Physics Reports by Nilles \cite{Nilles}, Nanopoulos 
\cite{Nanopoulos}, Haber and Kane \cite{Haber}, and Chung, Everett, Kane, King, 
Lykken and Wang \cite{Chung}. See also the TASI lectures of 
Dine \cite{Dine}, the Les Houches lectures of Ross \cite{Ross} and the review by Raby \cite{Raby}. 

For Kaluza-Klein theories and supergravity, see the Shelter Island lectures of Witten 
\cite{Wittenfermion}, the Physics Reports by
Duff, Nilsson and Pope \cite{Duffnilssonpopekaluza}, the reprint 
volume by Appelquist, Chodos and Freund \cite{Appelquist}, the books by 
Castellani, D'Auria and Fre \cite{Castellani} and Salam and Sezgin 
\cite{Salamsezgin} and the reviews by Duff \cite{Klein1,Klein2}.

\subsection{String theory}

To paraphrase Weinberg:

{\it
Supergravity is itself only an effective nonrenormalizable theory
which breaks down at the Planck energies.  So if there is any truth to
supersymmetry then any realistic theory must eventually be enlarged to
superstrings which are ultraviolet finite.  Supersymmetry without 
superstrings is not an option.}

Following the 1984 superstring revolution, the emphasis in the search
for a final theory shifted away from the spacetime aspects of
supergravity towards the two-dimensions of the string worldsheet.
The five consistent superstrings: Type I, Type IIA, Type IIB,
Heterotic $E_{8} \times E_{8}$ and Heterotic $SO(32)$ all feature spacetime local 
supersymmetry in ten dimensions.  It plays a crucial part in 
discussions of superstring compactification from ten to four 
dimensions and, inter alia, has also stimulated research in pure 
mathematics, for example Calabi-Yau manifolds and manifolds of 
exceptional holonomy.

Introductions to string theory may be found in the books by Green,
Schwarz and Witten \cite{GSW} and Polchinski \cite{Polchinskistrings}.

\subsection{M-theory}

To paraphrase Weinberg again:

{\it
Superstring theory is itself only a perturbative theory which breaks down
at strong coupling.  So if there is any truth to supersymmetry then any
realistic theory must eventually be enlarged to the non-perturbative
M-theory, a theory involving higher dimensional extended objects: the
super p-branes. Supersymmetry without M-theory is not an option.}

In 1995 it was realized that a non-perturbative unification of
the five consistent superstring theories is provided by M-theory, whose
low-energy limit is eleven-dimensional supergravity. In addition to
strings,   M-theory involves p-dimensional extended objects, namely the
p-branes which couple to the background fields of D=11 supergravity.  
This resolved the old mystery of why local supersymmetry allows 
a maximum of eleven dimensions while superstrings stop at ten.  
Indeed, many of the p-branes were first understood as classical 
solutions of the supergravity field equations.  As a result, supergravity has returned 
to center stage.

M-theory is regarded by many as the dreamed-of final theory and has
accordingly received an enormous amount of attention.  It is curious,
therefore, that two of the most basic questions of M-theory have until
now remained unanswered:

i) {\it What are the D=11 symmetries?}

In the section \ref{hidden} we will argue that the equations of M-theory 
possess previously unidentified hidden spacetime (timelike and null) 
symmetries in addition to the well-known hidden internal (spacelike) 
symmetries.  For $11 \geq d \geq 3$, these coincide with the generalized 
structure groups discussed below and take the form ${\cal 
G}=\SO(d-1,1) \times G(spacelike)$, ${\cal G}= \ISO(d-1) \times 
G(null)$ and ${\cal G}=\SO(d) \times G(timelike)$ with $1\leq d<11$.  
For example, $G(spacelike)=\SO(16)$, $G(null)=[\SU(8) \times 
\U(1)]\ltimes \R^{56}$ and $G(timelike)=\SO^*(16)$ when $d=3$.  The 
nomenclature derives from the fact that these symmetries also show up 
in the spacelike, null and timelike dimensional reductions of the 
theory.  However, we emphasize that we are proposing them as 
background-independent symmetries of the full unreduced and 
untruncated $D=11$ equations of motion, not merely their dimensional 
reduction.  Although extending spacetime symmetries, there is no 
conflict with the Coleman-Mandula theorem.  A more speculative idea is 
that there exists a yet-to-be-discovered version of $D=11$ 
supergravity or $M$-theory that displays even bigger hidden symmetries 
corresponding to ${\cal G}$ with $d\leq 3$ which 
could be as large as $SL(32,R)$.

ii) {\it How many supersymmetries can vacua of M-theory preserve?}

The equations of M-theory display the maximum number of supersymmetries
$N$=32, and so $n$, the number of supersymmetries preserved by a
particular vacuum, must be some integer between 0 and 32.  But are
some values of $n$ forbidden and, if so, which ones?  For quite some
time it was widely believed that, aside from the maximal $n=32$, $n$
is restricted to $0\leq n\leq 16$ with $n=16$ being realized by the
fundamental BPS objects of M-theory: the M2-brane, the M5-brane, the
M-wave and the M-monopole.  The subsequent discovery of intersecting
brane configurations with $n=0$, 1, 2, 3, 4, 5, 6, 8, 16 lent credence
to this argument.  On the other hand, it has been shown
that all values $0\leq n \leq 32$ are in principle allowed by the
M-theory algebra discussed in section \ref{algebra}, and examples of 
vacua with $16< n < 32$ have indeed 
since been found.  In fact, the values of $n$ that have been found 
``experimentally'' to date are: 
$n=$0,1,2,3,4,5,6,8,10,12,14,16,18,20,22,24,26,28,32.

In M-theory vacua with vanishing 4-form $F_{(4)}$, one can invoke the
ordinary Riemannian holonomy $H \subset \SO(10,1)$ to account for
unbroken supersymmetries $n=1, 2, 3, 4, 6, 8, 16, 32$.  To explain the
more exotic fractions of supersymmetry, in particular $16<n<32$, we
need to generalize the notion of holonomy to accommodate non-zero
$F_{(4)}$.
 In section \ref{counting} we show that the number of supersymmetries 
 preserved by an M-theory vacuum is given by the number of singlets 
 appearing in the decomposition of the 32-dimensional representation of 
 ${\cal G}$ under ${\cal G} \supset {\cal H}$ where ${\cal G}$ are 
 generalized structure groups that replace $SO(1,10)$ and ${\cal H}$ 
 are generalized holonomy groups.  In general we require the maximal 
 ${\cal G}$, namely $SL(32,R)$, but smaller ${\cal G}$ appear in 
 special cases such as product manifolds.

Reviews
of $M$-theory may be found in the paper by Schwarz \cite{Schwarzpower}, 
the paper by Duff \cite{M}, the book by Duff \cite{Duffworld1}, the 
lectures of Townsend \cite{TownsendM} and the books by Kaku
\cite{Kaku1,Kaku2}.  Reviews on supermembranes are given in the 
Physics reports of Duff, Khuri and Lu
\cite{Khuristring},the TASI lectures by Duff \cite{Duffsupermembranes5} 
and the papers by Duff \cite{Fifteen5,Classical5} and Stelle \cite{Stelle5}, 
the books by Polchinski \cite{Polchinskistrings}, Johnson 
\cite{Dbranes} and Ortin \cite{Ortin}.

\section{Simple supersymmetry in four dimensions}

\subsection{The algebra}

The $N=1$ supersymmetry algebra takes the form
\[
\{Q_\alpha,Q_\beta\}=2(\gamma_aC)_{\alpha\beta}P^\mu
\]
\[
[Q_\alpha,P_\mu]=0~~~~~~~~~
\]
\[
[Q_\alpha,J_{\mu\nu}]={1\over2}(\sigma_{\mu\nu})_\alpha^\beta Q_\beta
\]
\begin{equation}
[Q_\alpha,R]=i(\gamma_5)_\alpha^\beta Q_\beta
\label{n=1algebra}
\end{equation}
together with the commutation relations of the Poincare group. 

\subsection{Wess-Zumino model} 

The simplest representation of this algebra is provided by the 
Wess-Zumino multiplet which consists of 2 scalars $A$ and $B$, a 
4-component fermion $\chi$ and two auxiliary fields $F$ and $G$.
The free Wess-Zumino Lagrangian is given by
\[
{\cal L}_{WZ}¥=-\frac{1}{2}\left[(\partial_{\mu}A)^{2}+(\partial_{\mu}B)^{2}
+{\bar \chi}\gamma^{\mu}\partial_{\mu} \chi -F^{2}- G^{2}\right]
\]
The action is invariant under the supersymmetry transformations
\[
\delta A=\frac{1}{2} {\bar \epsilon} \chi\]
\[
\delta B=- \frac{1}{2} {\bar \epsilon} \gamma_{5} \chi
\]
\[
\delta 
\chi=\frac{1}{2}\left[\gamma^{\mu}\partial_{\mu}(A-i\gamma_{5}B)
+(F+i\gamma_{5}G)\right]\epsilon
\]
\[
\delta F=\frac{1}{2}{\bar \epsilon}\gamma^{\mu}\partial_{\mu} \chi
\]
\begin{equation}
\delta G=\frac{1}{2}{\bar \epsilon}\gamma_{5}\gamma^{\mu}\partial_{\mu} 
\chi
\end{equation}

It is now easy to see why supersymmetry is sometimes called ``the 
square root of a translation''. For example
\begin{equation}
[\delta_{1}, \delta_{2}]A=a^{\mu}\partial_{\mu} A
\end{equation}
where
\begin{equation}
a^{\mu}= {\bar \epsilon_{1}}\gamma^{\mu}\epsilon_{2}
\end{equation}

\subsection{Super Yang-Mills}

Another representation is provided by the vector multiplet which 
consists of a set of vectors $A_{\mu}^{i}$, fermions $\lambda^{i}$ and  
auxiliary fields $D^{i}$. The Yang-Mills Lagrangian is given by
\begin{equation}
{\cal L}_{YM}=-{1\over4}(F_{\mu\nu}^i)^2
-{1\over2}\bar\lambda^i{\cal
D}\!\!\!/\lambda^i+{1\over2}(D^i)^2
\end{equation}
The action is invariant under the supersymmetry transformations
\[
\delta A_{\mu}^i=\bar\varepsilon\gamma_{\mu}\lambda^i
\]
\[
\delta\lambda^i=\left(-{1\over2}\sigma^{\mu\nu}F_{\mu\nu}^i
+i\gamma_5D^i\right)\varepsilon
\]
\begin{equation}
\delta D^i=i\bar\varepsilon\gamma_5{\cal D}\!\!\!/\lambda^i
\end{equation}
where
\begin{equation}
F_{\mu\nu}^i=\partial_\mu A_{\nu}^i-\partial_\nu A_{\mu}^i- gc_{jk}^{i} 
A_{\mu}^j A_{\nu}^k
\end{equation}

\subsection{Simple supergravity}

Finally we come to the tensor multiplet consisting of a vierbein $e_a^{\ 
\mu}$, a gravitino $\psi_\mu$ and auxiliary fields $b_{\mu}$, $M$ 
and $N$. The supergravity lagrangian is
\begin{equation}
{\cal L}_{SUGRA}¥={e\over2\kappa^2}R
-{1\over2}\bar\psi_\mu R^\mu
-{1\over3}e(M^2+N^2-b_\mu b^\mu)
\end{equation}
where
\begin{equation}
R=R_{\mu\nu}^{\ \ ab}e_a^{\ \mu}e_b^{\ \nu}
\end{equation}
and
\begin{equation}
\frac{1}{4}R_{\mu\nu}^{\ \ ab}{\sigma_{ab}}=[D_\mu,D_\nu]
\end{equation}
The transformations are now those of local supersymmetry where 
$\epsilon=\epsilon(x):$
\[
\delta e^{\ a}_\mu=\kappa\bar\varepsilon\gamma^a\psi_\mu
\]
\[
\delta\psi_\mu=2\kappa^{-1}D_\mu\big(w(e,\psi)\big)\varepsilon
+i\gamma_5\left(b_\mu-{1\over3}\gamma_\mu/\!\!\!b\right)\varepsilon
-{1\over3}\gamma_\mu(M+i\gamma_5N)\varepsilon
\]
\[
\delta M=-{1\over2}e^{-1}\bar\varepsilon\gamma_\mu
R^\mu-{\kappa\over2}i\bar\varepsilon\gamma_5\psi_\nu
b^\nu-\kappa\bar\varepsilon
\gamma^\nu\psi_\nu
M+{\kappa\over2}\bar\varepsilon(M+i\gamma_5N)\gamma^\mu\psi_\mu
\]
\[
\delta N=-{e^{-1}\over2}i\bar\varepsilon\gamma_5\gamma_\mu
R^\mu+{\kappa\over2}\bar\varepsilon\psi_\nu b^\nu-\kappa\bar\varepsilon
\gamma^\nu\psi_\nu
N-{\kappa\over2}i\bar\varepsilon\gamma_5(M+i\gamma_5N)
\gamma^\mu\psi_\mu
\]
\begin{equation}
\delta b_\mu={3i\over2}e^{-1}\bar\varepsilon\gamma_5\left(g_{\mu\nu}-{1\over3}
\gamma_\mu\gamma_\nu\right)R^\nu+\kappa\bar\varepsilon
\gamma^\nu
b_\nu\psi_\mu-{\kappa\over2}\bar\varepsilon\gamma^\nu\psi_\nu b_\mu
\quad-{\kappa\over2}i\bar\psi_\mu\gamma_5(M+i\gamma_5N)
\varepsilon-{i\kappa\over4}\varepsilon_\mu^{\ bcd}b_b\bar\varepsilon\gamma_5
\gamma_c\psi_d
\end{equation}
where
\begin{equation}
R^\mu=\varepsilon^{\mu\nu\rho\kappa}i\gamma_5\gamma_\nu
D_\rho\big(w(e,\psi)\big)\psi_\kappa
\end{equation}
\begin{equation}
D_\mu\big(w(e,\psi)\big)=\partial_\mu+\frac{1}{4}w_{\mu
ab}{\sigma^{ab}}
\end{equation} 
and
\[
w_{\mu ab}={1\over2}e^\nu_{\ a}(\partial_\mu e_{b\nu}-\partial_\nu
e_{b\mu})-{1\over2}e_b^{\ \nu}(\partial_\mu e_{a\nu}-\partial_\nu
e_{a\mu})
\]
\[
\quad-{1\over2}e_a^{\ \rho}e_b^{\ \sigma}(\partial_\rho e_{\sigma
c}-\partial_\sigma a_{\rho c})e_\mu^{\ c}
\]
\begin{equation}
\quad+{\kappa^2\over4}(\bar\psi_\mu\gamma_a\psi_b
+\bar\psi_a\gamma_\mu\psi_b-\bar\psi_\mu\gamma_b\psi_a)
\end{equation}

\subsection{Off-shell versus on-shell}

Since the auxiliary fields $F$, $G$, $D^{i}$, $b_{\mu}$, $M$ 
and $N$ enter only algebraically in the 
Lagrangians, they may be eliminated by their equations of motion, if so desired. 
With the auxiliary fields, however, the algebra closes off-shell 
whereas it closes only on-shell without them. It is useful to count the 
number of degrees of freedom in both cases.
 
Off-shell: For the Wess-Zumino multiplet, $A$ 
and $B$ each count 1, $\chi$ counts 4 and $F$ and $G$ each count 1, 
making 4 bose and 4 fermi in total. For the vector multiplet, $A_{\mu}$ counts 
3, $\lambda$ counts 4 and $D$ counts 1, making 4 bose and 4 fermi in 
total. For the supergravity multiplet, $e^\mu_{\ a}$ counts 
$16-10=6$, $\psi_\mu$ counts $16-4=12$, $b_{\mu}$ counts 4 and $M$ and 
$N$ each count 1, making 12 bose plus 12 fermi in total.

On-shell: For the Wess-Zumino multiplet, $A$ 
and $B$ each count 1, $\chi$ counts 2 and $F$ and $G$ each count 0, 
making 2 bose and 2 fermi in total. For the vector multiplet, $A_{\mu}$ counts 
2, $\lambda$ counts 2 and $D$ counts 0, making 2 bose and 2 fermi in 
total. For the supergravity multiplet, $e^\mu_{\ a}$ counts 
2, $\psi_\mu$ counts 2, $b_{\mu}$ counts 0 and $M$ and 
$N$ each count 0, making 2 bose plus 2 fermi in total. 

Note that supersymmetry always requires equal number of bose and fermi 
degrees of freedom both off-shell and on-shell. 

\subsection{Particle phenomenology}

The requirement of chirality limits us to $N = 1$ and the most general such
theory consists of $N = 1$ supergravity coupled to $N = 1$ Yang-Mills and
$N = 1$ chiral multiplets. This theory is characterized by three functions
of the chiral multiplets: the superpotential $W$, the K{a}hler potential
$K$ and the gauge function $f$. The function $f$ is real while $W$ and $K$ are holomorphic.

Within this framework, one might wish to embed the standard model gauge
groups $SU(3) \times SU(2) \times U(1)$ and three families of quarks and
leptons. Of course this immediately doubles the number of elementary
particles, since every particle we know of acquires a superpartner,
none of which can be identified with a known particle. These have
names like gauginos (winos, zinos, photinos and gluinos), higgsinos,
squarks and sleptons. Moreover, unbroken supersymmetry implies that
these superpartners are degenerate in mass with the known particles in
obvious disagreement with experiment. In any realistic theory, therefore,
supersymmetry must be broken.  Since the equations of motion of the
only known quantum consistent theories of gravity are supersymmetric,
this breaking must be spontaneous.  However, the resulting low-energy
theory can be represented by a globally supersymmetric Lagrangian
${\cal L}_{soft}$ with explicit but soft breaking terms. By soft we
mean operators of dimensions 2 or 3. The bottom-up approach is thus to
write down such a minimal supersymmetric standard model (MSSM) with
mass parameters that are typically of the order of the electroweak
to TeV scale. Counting the masses, coupling constants and phases, the
most general such model has 124 parameters. Of course, experiment can
provide constraints. Its claimed successes include resolutions of: the
technical gauge hierarchy problem, the electroweak symmetry breaking
problem, the gauge coupling unification problem, the cold dark matter
problem and the baryon asymmetry problem.

In the literature, there is a plethora of different top-down proposals
for how this spontaneous supersymmetry breaking may come about. The
obvious tree-level TeV breaking in which either the F or D auxiliary
fields acquire vacuum expectation values seems to be ruled out by
experiment. One alternative is the hidden sector framework where the
theory can be split into two sectors with no renormalizable couplings
between them: an observable sector containing the SM model particles and
their superpartners, and hidden sector in which supersymmetry is broken
by a dynamical mechanism such as gaugino condensation. The scale of
supersymmetry breaking $M_S$ is hierarchically higher than a TeV.

There are various versions of these hidden sector models: gravity
mediated models, gauge mediated models, bulk mediated models. In the
latter scenario, the observable and hidden sectors reside on different
branes embedded in a bulk spacetime of higher dimension.

Another alternative is D-term breaking which arises in extensions of the MSSM
to GUTs or strings.

The hope, of course, is that the correct mechanism will be selected by
the fundamental theory but owing to the vacuum degeneracy problem, there
has been very little progress in this respect. In fact, neither string
theory nor M-theory has yet been able to fix any of the 124 parameters.

\section{Extended supersymmetry}

\subsection{The algebra}
   
To discuss extended supersymmetry, it is more convenient to rewrite 
the (anti)commutation relations (\ref{n=1algebra}) in terms of 
two-component Weyl spinors $Q_{\alpha}$ and ${\bar Q}_{\dot \alpha}$
\[
\{Q_{\alpha},Q_{\beta}\}=\{{\bar Q}_{\dot \alpha},{\bar Q}_{\dot 
\beta}\}=0
\]
\[
\{{Q}_{\alpha},{\bar Q}_{\dot \beta}\}=2\sigma^{\mu}_{\alpha \dot 
\beta}P_{\mu}
\]
\begin{equation}
[Q_{\alpha},P_{\mu}]= [{\bar Q}_{\dot \alpha},P_{\mu}]=0 
\label{simple}
\end{equation}
in which dotted and undotted indices take the values $\alpha, \dot 
\alpha=1,2$.

We now allow for a set of $Q_{\alpha}$, labelled by an index $L$, 
which transform according to some representation of a compact Lie 
group $G$, and ${\bar Q}_{\dot \alpha}^{L}=(Q_{\alpha}^{L})^{*}$ which 
transform according to the complex conjugate representation. The simple 
supersymmetry algebra (\ref{simple}) now generalizes to the extended 
supersymmetry algebra
\[
\{Q_{\alpha}^{L},Q_{\beta}^{M}\}=\epsilon_{\alpha\beta}Z^{LM}
\]
\[
\{{\bar Q}_{\dot \alpha}^{L},{\bar Q}_{\dot 
\beta}^{M}\}=\epsilon_{\dot \alpha \dot \beta}Z^{LM}
\]
\[
\{{Q}_{\alpha}^{L},{\bar Q}_{\dot \beta}^{M}\}=2\delta^{LM}\sigma^{\mu}_{\alpha \dot 
\beta}P_{\mu}
\]
\[
[Q_{\alpha}^{L},P_{\mu}]= [{\bar Q}_{\dot \alpha}^{L},P_{\mu}]=0 
\]
\[
[{Q}_{\alpha}^{L},B_{l}]=iS_{l}^{LM}Q_{\alpha}^{M}
\]
\begin{equation}
[B_{l},B_{m}]=if_{lmk}B_{k}
\label{extended}
\end{equation}
where $S_{l}^{LM}$ are the hermitian matrices of the representation  
containing the ${Q}_{\alpha}^{L}$ and the $B_{k}$ are the generators of the internal 
symmetry group $G$. The $Z^{LM}$ are central charges which commute 
with all the other generators. 

\subsection{Multiplets}

We shall not detail the representation theory of the extended 
supersymmetry 
algebra (\ref{extended}) but simply quote some results. Massless irreducible 
representations with maximum helicity 1 and 2 are 
tabulated in Tables \ref{1} and \ref{2}, respectively.
Some massive representations with and without central charges are 
tabulated in Tables \ref{3} and \ref{4}.

Discussions of representations of extended supersymmetry may be found in the Trieste 
Lectures of Ferrara and Savoy \cite{Savoy} and in the review of 
Strathdee \cite{Strathdee}.

\begin{table}
 $$\offinterlineskip\halign{\hfil$# 
$\hfil&\quad\strut#&\vrule\quad \hfil$# $\hfil&\quad\hfil$# 
$\hfil&\quad\hfil$# $\hfil&\quad\hfil$# $\hfil&\quad# \cr 
\noalign{\hrule} \qquad N&&&&&&\cr {\rm Spin}&&1&1&2&2&4\cr 
\noalign{\hrule} {\rm Spin}\ 1&&-&1&1&-&1\cr {\rm Spin}\ 
{1\over2}&&1&1&2&2&4\cr {\rm Spin}\ 0&&2&-&2&4&6\cr
\noalign{\hrule}}$$
\par
\caption{Multiplicities for massless irreducible representations with 
maximal helicity 1 or less}
\label{1}
\end{table}

\begin{table} 
$$\offinterlineskip\halign{\hfil$# $\hfil&\quad\strut#&\vrule\quad 
\hfil$# $\hfil&\quad\hfil$# $\hfil&\quad\hfil$# $\hfil&\quad\hfil$# 
$\hfil& \quad\hfil$# $\hfil&\quad\hfil$# $\hfil&\quad\hfil$# 
$\hfil&\quad# \cr \noalign{\hrule} \qquad N&&&&&&&&&\cr {\rm 
Spin}&&1&2&3&4&5&6&7&8\cr \noalign{\hrule} {\rm Spin}\ 
2&&1&1&1&1&1&1&1&1\cr {\rm Spin}\ {3\over2}&&1&2&3&4&5&6&8&8\cr {\rm 
Spin}\ 1&&&1&3&6&10&16&28&28\cr {\rm Spin}\ 
{1\over2}&&&&1&4&11&26&56&56\cr {\rm Spin}\ 0&&&&&2&10&30&70&70\cr
\noalign{\hrule}}$$
\caption{Multiplicity for massless on-shell representations with maximal helicity 2.}
\label{2}
\end{table}

\subsection{Auxiliary fields?}

When we come to extended supersymmetry and higher dimensions, the 
off-shell formalism is not always available.  In $D=4$, the finite set 
of auxiliary fields has been worked out only for $N=1$ and $N=2$ 
multiplets and some $N=4$ supergravity/matter combinations.  No theory 
beyond half-maximal has an off-shell formulation with a finite number 
of auxiliary fields.  Harmonic superspace can extend the range but at 
the price of an infinite number.  There is no known off-shell 
formulation for the maximally supersymmetric theories.  This is a 
drawback since non-renormalization theorems are most transparent in 
the off-shell formalism.  For example, the finiteness of the maximally 
supersymmetric N=4 Yang-Mills theory leads one to wonder whether the 
maximally supersymmetric N=8 supergravity might also have some peculiar 
ultraviolet properties. 

The absence of a complete off-shell formalism also remains something of a mystery: is 
there some deeper meaning to all this?

\begin{table}
$$\offinterlineskip\halign{\hfil$# $\hfil&\quad\strut#&\vrule\quad
\hfil$# $\hfil&\quad\hfil$# $\hfil&\quad\hfil$# $\hfil&\quad\hfil$#
$\hfil&\quad\strut#&\vrule\quad\hfil$# $\hfil&\quad\hfil$#
$\hfil&\quad\hfil$# $\hfil
&\quad\strut#&\vrule\quad\hfil$# $\hfil&\quad\hfil$#
$\hfil&\quad\strut#&\vrule
\quad# \cr
\noalign{\hrule}
\qquad N&&&&&&&&&&&&&&\cr
{\rm Spin}&&&1&&&&&2&&&&3&&4\cr
\noalign{\hrule}
{\rm Spin}\ 2&&&&&1&&&&1&&&1&&1\cr
{\rm Spin}\ {3\over2}&&&&1&2&&&1&4&&1&6&&8\cr
{\rm Spin}\ 1&&&1&2&1&&1&4&5+1&&6&14+1&&27\cr
{\rm Spin}\ {1\over2}&&1&2&1&&&4&5+1&4&&14&14^\prime+6&&48\cr
{\rm Spin}\ 0&&2&1&&&&5&4&1&&14^\prime&14&&42\cr
\noalign{\hrule}}$$
\par
\caption{Some massive representations (without central charges) labelled in 
terms of the $USp(2N)$ representations.}
\label{3}
\end{table}

\begin{table}
$$\offinterlineskip\halign{\hfil$# $\hfil&\quad\strut#&\vrule\quad
\hfil$# $\hfil&\quad\hfil$# $\hfil&\quad\strut#&\vrule\quad\hfil$#
$\hfil&
\quad\hfil$# $\hfil&\quad\strut#&\vrule\quad\hfil$#
$\hfil&\quad\hfil$# $\hfil
&\quad\strut#&\vrule\quad# \cr
\noalign{\hrule}
\qquad N&&&&&&&&&&&\cr
{\rm Spin}&&2&&&4&&&6&&&8\cr
\noalign{\hrule}
{\rm Spin}\ 2&&&&&&&&&1&&1\cr
{\rm Spin}\ {3\over2}&&&&&&1&&1&6&&8\cr
{\rm Spin}\ 1&&&1&&1&4&&6&14+1&&27\cr
{\rm Spin}\ {1\over2}&&1&2&&4&5+1&&14&14^\prime+6&&48\cr
{\rm Spin}\ 0&&2&1&&5&4&&14^\prime&14&&42\cr
\noalign{\hrule}}$$
\par
\caption{Some massive representations with one central charge 
$(|Z|=m)$.  All states are complex.}
\label{4}
\end{table}

Early discussions of ultraviolet divergences in extended supergravity may 
be found in the Trieste Lectures by Duff \cite{ultra} and the paper 
by Howe and Lindstrom \cite{Lindstrom}, and up-to-date ones in the 
review by Bern at al \cite{Bern} and the paper by Howe and 
Stelle \cite{Howestelle}.

\section{Eleven dimensions}

\subsection{The algebra}
\label{algebra}
Eleven is the maximum spacetime dimension in which one can formulate
a consistent supersymmetric theory, as was first recognized by Nahm 
in his classification of supersymmetry algebras.  The easiest way
to see this is to start in four dimensions and note that one 
supersymmetry
relates states differing by one half unit of helicity. If we now make 
the
reasonable assumption that there be no massless particles with spins 
greater than
two, then we can allow up to a maximum of $N=8$ supersymmetries 
taking us from
helicity $-2$ through to helicity $+2$.  Since the  minimal 
supersymmetry
generator is a Majorana spinor with four off-shell components, this 
means a total
of $32$ spinor components. Now in a spacetime with $D$ dimensions and
signature $(1,D-1)$, the maximum value of $D$ admitting a $32$ 
component
spinor is $D=11$. (Going to $D=12$, for example, would require $64$ 
components.)
See Table \ref{minimal}\footnote{Conventions differ on how to count 
the supersymmetries and the more usual conventions are that $N_{max}=8$ 
in $D=5$ and $N_{max}=4$ in $D=7$ }.  Furthermore, $D=11$ emerges 
naturally as the maximum dimension admitting supersymmetric extended 
objects. 

The full D=11 supertranslation algebra is
\begin{equation}
\{Q_\alpha,Q_\beta\} = (C\Gamma^M)_{\alpha\beta} P_M + 
(C\Gamma_{MN})_{\alpha\beta} Z_{}^{MN} + 
(C\Gamma_{MNPQR})_{\alpha\beta}
Z_{}^{MNPQR}\, .
\label{eq:memftot}
\end{equation}
Note that the total number of algebraically independent charges that 
could
appear
on the right hand side is 528. The number actually appearing is
\begin{equation}
11+55+462 =528
\label{count}
\end{equation}
so the algebra (\ref{eq:memftot}) is `maximally extended'. The three 
types of
charge appearing on the right hand side are those associated with the
supergraviton, the supermembrane and the superfivebrane, which are 
the three
basic ingredients of $M$-theory. The time components $Z_{0I}$ and 
$Z_{0IJKL}$ are associated with the 8-brane and 6-brane of Type IIA 
theory that arise on compactification to D=10.
 
The M-theory algebra is treated in the 
papers by Townsend\cite{Townsend} and Gauntlett and Hull 
\cite{Gauntletthull1}.  

\begin{table}
\halign{\indent #&\qquad\hfil# \hfil&\quad\hfil
#\hfil&\quad\hfil # \hfil &
\quad \hfil # \hfil &\quad \hfil # \hfil &\quad #\hfil\cr
&&&Dimension & Minimal Spinor& Supersymmetry&\cr
&&&($D$ or $d$) & ($M$ or $m$) & ($N$ or $n$)&\cr
&&&11 & 32 & 1&\cr
&&&10 & 16 & 2, 1&\cr
&&&9 & 16 & 2, 1&\cr
&&&8 & 16 &2, 1&\cr
&&&7 & 16 & 2, 1&\cr
&&&6 & 8 & 4, 3, 2, 1&\cr
&&&5 & 8 & 4, 3, 2, 1&\cr
&&&4 & 4 & 8, $\ldots$, 1&\cr
&&&3 & 2 & 16, $\ldots$, 1&\cr
&&&2 & 1 & 32, $\ldots$, 1&\cr}
\caption{Minimal spinor components and supersymmetries.}
\label{minimal}
\end{table}

\subsection{The multiplet}
\label{Content}

Not long after Nahm's paper, Cremmer, Julia and Scherk 
realized that supergravity not only permits up to seven extra 
dimensions but in fact takes its simplest and most elegant form when 
written in its full eleven-dimensional glory.  The unique $D=11,N=1$ 
supermultiplet is comprised of a graviton $g_{MN}$, a gravitino 
$\psi_M$ and $3$-form gauge field $A_{MNP}$ with $44$, $128$ and $84$ 
physical degrees of freedom, respectively.  For a counting of on-shell 
degrees of freedom in higher dimensions, see Table \ref{Degrees}.
\begin{table}
\[
\begin{array}{llll}
&d-bein&e_{M}¥{}^{A}¥&D(D-3)/2\\
&gravitino&\Psi_{M}¥&2^{(\alpha-1)}¥(D-3)\\
&p-form&A_{M_{1}¥M_{2}¥\ldots M_{p}¥}¥&
\left(
\begin{array}{c}
D-2\\
p
\end{array}
\right)
\\
&spinor&\chi&2^{(\alpha-1)}¥
\end{array}
\]
\caption{On-shell degrees of freedom in $D$ dimensions. $\alpha=D/2$ 
if $D$ is even, 
$\alpha=(D-1)/2$ if $D$ is odd. We assume Majorana fermions and 
divide by 
two if the fermion is Majorana-Weyl. Similarly, we assume real 
bosons and divide by two if the tensor field strength is self-dual. } 
\label{Degrees}
\end{table}
The theory may also be formulated in 
superspace.  Ironically,
however, these extra dimensions were not at first taken seriously but 
rather
regarded merely as a useful device for deriving supergravities in four
dimensions. Indeed $D=4,N=8$ supergravity was first obtained by 
Cremmer and Julia
 via the process of {\it dimensional reduction} i.e. by
requiring that all the fields of $D=11,N=1$ supergravity be 
independent of the
extra seven coordinates. 

\subsection{D=11 supergravity}

 For future reference we record the bosonic field
equations
\begin{equation}
R_{MN}=\frac{1}{12}\left(F_{MPQR}F_{N}{}^{PQR}-\frac{1}{12}g_{MN}
F^{PQRS}F_{PQRS}\right)
\end{equation}
and
\begin{equation}
d*\!F_{(4)}+\fft12F_{(4)} \wedge F_{(4)}=0,
\end{equation}
where $F_{(4)}=dA_{(3)}$.  The supersymmetry transformation rule of the
gravitino reduces in a purely bosonic background to
\begin{equation}
\delta \Psi_{M}={\cal D}_{M} \epsilon,
\label{eq:11gto}
\end{equation}
where the parameter $\epsilon$ is a 32-component anticommuting spinor,
and where
\begin{equation}
{\cal D}_{M}=D_{M}-
\frac{1}{288}(\Gamma_M{}^{NPQR}-8\delta_M^N\Gamma^{PQR})F_{NPQR},
\label{supercovariant}
\end{equation}
where $\Gamma^{A}$ are the $D=11$ Dirac matrices and 
$\Gamma_{AB}=\Gamma_{[A}\Gamma_{B]}$.  Here $D_{M}$ is the usual 
Riemannian covariant derivative involving the connection $\omega_{M}$ 
of the usual structure group $\Spin(10,1)$, the double cover of 
$\SO(10,1)$, %
\begin{equation}
D_{M}=\partial_{M}+\frac{1}{4}\omega_{M}{}^{AB}\Gamma_{AB}
\label{usual}
\end{equation} 

For many years the Kaluza-Klein idea of taking extra dimensions 
seriously was largely
forgotten but the arrival of eleven-dimensional supergravity provided 
the missing
impetus.  
The kind of four-dimensional world we end up with depends on how we
{\it compactify} these extra dimensions: maybe seven of them would 
allow us to
give a gravitational origin, {a} la Kaluza-Klein, to the strong and 
weak forces as
well as the electromagnetic.  In a very influential paper, Witten 
 drew attention to the fact that in such a scheme the
four-dimensional gauge group is determined by the {\it isometry} 
group of the
compact manifold ${\cal K}$. Moreover, he proved (what to this day 
seems 
to be merely a gigantic coincidence) that seven is not only the 
maximum dimension
of ${\cal K}$  permitted by supersymmetry but the minimum needed for 
the isometry
group to coincide with the standard model gauge group $SU(3) \times 
SU(2)
\times U(1)$.

In the early 80's there was great interest in 
four-dimensional $N$-extended supergravities for which the global 
$SO(N)$ is 
promoted to a gauge
symmetry. In these theories the underlying supersymmetry algebra is 
no longer 
Poincar{e} but rather anti-de Sitter ($AdS_4$) and the
Lagrangian has a non-vanishing cosmological constant $\Lambda$ 
proportional to
the square of the gauge coupling constant $g$:  
\begin{equation}
 G\Lambda \sim -g^{2}¥ \label{G} 
 \end{equation} where $G$ is Newton's constant.  The $N>4$ gauged 
 supergravities were particularly interesting since the cosmological 
 constant $\Lambda$ does not get renormalized and hence the $SO(N)$ 
 gauge symmetry has vanishing $\beta$-function\footnote{For $N\leq 4$, 
 the $beta$ function (which receives a contribution from the spin $3/2$ 
 gravitinos) is positive and the pure supergravity theories are not 
 asymptotically free.  The addition of matter supermultiplets only 
 makes the $\beta$ function more positive and hence gravitinos can 
 never be confined.}.  The relation 
 (\ref{G}) suggested that there might be a Kaluza-Klein interpretation 
 since in such theories the coupling constant of the gauge group 
 arising from the isometries of the extra dimensions is given by 
 \begin{equation} g^{2}¥ \sim Gm^{2}¥ \label{e} \end{equation} where 
 $m^{-1}¥$ is the size of the compact space.  Moreover, there is 
 typically a negative cosmological constant \begin{equation} \Lambda\sim 
 -m^{2}¥ \label{Lambda} \end{equation} Combining (\ref{e}) and 
 (\ref{Lambda}), we recover (\ref{G}).  Indeed, the maximal $(D=4,N=8)$ 
 gauged supergravity was seen to correspond to the massless sector of 
 $(D=11,N=1)$ supergravity compactified on an $S^7$ whose metric admits 
 an $SO(8)$ isometry and $8$ Killing spinors.  An important ingredient 
 in these developments that had been insufficiently emphasized in 
 earlier work on Kaluza-Klein theory was that the $AdS_4 \times S^7$ 
 geometry was not fed in by hand but resulted from a {\it spontaneous 
 compactification}, i.e.  the vacuum state was obtained by finding a 
 stable solution of the higher-dimensional field equations.  The 
 mechanism of spontaneous compactification appropriate to the $AdS_4 
 \times S^7$ solution of eleven-dimensional supergravity was provided 
 by the Freund-Rubin mechanism in which the $4$-form field strength in 
 spacetime $F_{\mu\nu\rho\sigma}$ ($\mu=0,1,2,3$) is proportional to 
 the alternating symbol $\epsilon_{\mu\nu\rho\sigma}$: \begin{equation} 
 F_{\mu\nu\rho\sigma}¥ \sim \epsilon_{\mu\nu\rho\sigma} \label{F_{4}¥} 
 \end{equation}  By applying a 
 similar mechanism to the $7$-form dual of this field strength one 
 could also find compactifications on $AdS_{7} \times S^{4}$ whose 
 massless sector describes gauged maximal $N=4$, $SO(5)$ supergravity 
 in $D=7$.  Type IIB supergravity in $D=10$, with its self-dual 
 $5$-form field strength, also admits a Freund-Rubin compactification 
 on $AdS_{5}\times S^{5}$ whose massless sector describes gauged 
 maximal $N=8$ supergravity in $D=5$.

\begin{table}
\begin{center}
\begin{tabular}{ccc}
{\bf Compactification}&{\bf Supergroup}&{\bf Bosonic~subgroup}\\
$AdS_{4}¥\times S^{7}$¥&$OSp(4|8)$&$SO(3,2) \times SO(8)$\\
$AdS_{5}¥\times S^{5}$¥&$SU(2,2|4)$&$SO(4,2) \times SO(6)$\\
$AdS_{7}¥\times S^{4}$¥&$OSp(6,2|4)$&$SO(6,2) \times SO(5)$
\end{tabular}
\end{center}
\medskip
\caption{Compactifications and their symmetries.}
\label{supergroups}
\end{table}

In the three cases given above, the symmetry of the vacuum is
described by the supergroups $OSp(4|8)$, $SU(2,2|4)$ and $OSp(6,2|4)$
for the $S^7$, $S^5$ and $S^4$ compactifications respectively, as 
shown in Table \ref{supergroups}.  Each of these groups is known to 
admit 
the so-called singleton, doubleton or 
tripleton\footnote{Our nomenclature is based on the 
$AdS_{4}¥,AdS_{5}¥$ and $AdS_{7}¥$ groups having ranks $2,3$ and $4$, 
respectively,  and differs from that of G{u}naydin.} 
supermultiplets as shown in Table \ref{fields}. 
\begin{table}
\begin{center}
\begin{tabular}{ccc}
{\bf Supergroup}&{\bf Supermultiplet}&{\bf Field~content}\\
$OSp(4|8)$&$(n=8,d=3)$~singleton&8~scalars,8~spinors\\
$SU(2,2|4)$&$(n=4,d=4)$~doubleton&1~vector,8~spinors,6~scalars\\
$OSp(6,2|4)$&$((n_{+}¥,n_{-}¥)=(2,0),d=6)$
~tripleton&1~chiral~2-form,8~spinors,5~scalars
\end{tabular}
\end{center}
\medskip
\caption{Superconformal groups and their singleton, doubleton 
and 
tripleton representations.}
\label{fields}
\end{table}
We recall that 
singletons are those strange representations of $AdS$ first 
identified 
by Dirac which admit no analogue in flat spacetime. They 
have been much studied by Fronsdal and collaborators.

This Kaluza-Klein approach to $D=11$ supergravity
eventually fell out of favor for three reasons.  First, in spite of 
its maximal supersymmetry and other intriguing features, eleven 
dimensional supergravity was, after all, still a {\it field theory} of 
gravity with all the attendant problems of non-renormalizability.  
The resolution of this problem had to await the dawn of $M$-theory, since we 
now regard $D=11$ supergravity not as a fundamental theory in its own 
right but the effective low-energy Lagrangian of $M$-theory.
Second, as emphasized by Witten, it is impossible to derive by the 
conventional Kaluza-Klein technique of compactifying on a manifold a 
{\it chiral theory} in four spacetime dimensions starting from a 
non-chiral theory such as eleven-dimensional supergravity.  
Ironically, Horava and Witten were to solve this problem years later 
by compactifying $M$-theory on something that is not a manifold, namely $S^{1}/Z_{2}$.  
Thirdly, these $AdS$ vacua necessarily have non-vanishing 
cosmological constant unless cancelled by fermion condensates and this 
was deemed unacceptable at the time.  However, $AdS$ is currently 
undergoing a renaissance thanks to the $AdS/CFT$ correspondence.

 A discussion of spinors and Dirac matrices in $D$ spacetime dimensions may 
 be found in the reprint volume of Salam and Sezgin \cite{Salamsezgin} and 
 the book by West \cite{West}.  $D=11$ supergravity is discussed in 
 the paper of Cremmer, Julia and Scherk 
 \cite{Cremmerjuliascherk}.  A summary of the $S^{7}¥$ and other $X^{7}¥$ 
 compactifications of $D=11$ supergravity down to $AdS_{4}$ may be 
 found in the Physics Report of Duff, Nilsson and Pope 
 \cite{Duffnilssonpopekaluza}.
 
Discussions of anti-de Sitter space and singletons in supergravity may be found in the 
Physics Reports by Duff, Nilsson and Pope 
\cite{Duffnilssonpopekaluza}, the
review by G{u}naydin in proceedings of the 1989 Trieste supermembrane conference 
\cite{Gunaydin1}, the book by Salam and Sezgin \cite{Salamsezgin}, and the 
TASI lectures by Duff \cite{Duff:1999rk}.

A review of the AdS/CFT correspondence may be found in 
Physics Reports of Aharony, Gubser, Maldacena, Ooguri and Oz 
\cite{Aharony:1999ti} and the TASI lectures of Maldacena 
\cite{Maldacenatasi}.


\section{Hidden spacetime symmetries in D=11}
\label{hidden}

\subsection{Spacelike, null and timelike reductions}
\label{sec:dimred}

Long ago, Cremmer and Julia pointed out that, when dimensionally 
reduced to $d$ dimensions, $D=11$ supergravity exhibits hidden 
symmetries.  For example ${\rm E}_7(global)\times \SU(8)(local)$ when 
$d=4$ and ${\rm E}_8(global) \times \SO(16)(local)$ when $d=3$.  
Cremmer and Julia concentrated on the case where all extra dimensions 
are spacelike.  Here we shall consider timelike and null reductions as 
well.  The global symmetries remain the same but we shall focus on the 
local symmetries.

In fact, in anticipation of applications to vacuum
supersymmetries in section \ref{counting}, we shall focus
particularly on the supercovariant derivative (\ref{supercovariant}) as
it appears in the gravitino variation of the dimensionally reduced
theory.  One finds that, after making a $d/(11-d)$ split, the Lorentz
subgroup $G=\SO(d-1,1) \times \SO(11-d)$ can be enlarged to the
generalized structure groups ${\cal G}=\SO(d-1,1) \times
G(spacelike)$, ${\cal G}= \ISO(d-1) \times G(null)$ and ${\cal G}=
\SO(d) \times G(timelike)$ arising in the spacelike, null and timelike
dimensional reduction, respectively.  As we shall see, these
generalized structure groups are the same as the hidden symmetries for
$d\geq3$ but differ for $d<3$.

First we consider a spacelike dimensional reduction corresponding to a
$d/(11-d)$ split.  Turning on only $d$-dimensional scalars, the
reduction ansatz is particularly simple
\begin{equation}
g^{(11)}_{MN}=\pmatrix{\Delta^{-1/(d-2)}g_{\mu\nu}&0\cr0&g_{ij}},\qquad
A^{(11)}_{ijk}=\phi_{ijk},
\end{equation}
where $\Delta=\det{g_{ij}}$.  For $d\le5$, we must also consider the
possibility of dualizing either $F_{(4)}$ components or (for $d=3$)
Kaluza-Klein vectors to scalars.  We will return to such possibilities
below.  But for now we focus on $d\ge6$.  In this case, a standard
dimensional reduction of the $D=11$ gravitino transformation 
(\ref{eq:11gto}) yields the $d$-dimensional gravitino transformation %

\begin{equation}
\delta\psi_\mu={\hat D}_\mu\epsilon
\label{psi}
\end{equation}
where
\begin{equation}
{\hat D}_\mu=\partial_\mu+\omega_\mu{}^{\alpha\beta}\gamma_{\alpha\beta} 
+Q_\mu{}^{ab}\Gamma_{ab}+\ft1{3!}e^{ia}e^{jb}e^{kc}\partial_\mu\phi_{ijk} 
\Gamma_{abc}.
\label{hatD}
\end{equation}
Here $\gamma_\alpha$ are $\SO(d-1,1)$ Dirac matrices, while $\Gamma_a$
are $\SO(11-d)$ Dirac matrices. For completeness, we also note that the $d$-dimensional dilatinos
transform according to
\begin{equation}
\delta\lambda_i=-\ft12\gamma^\mu[P_{\mu\,ij}\Gamma^j
-\ft1{36}(\Gamma_i{}^{jkl}-6\delta_i^j\Gamma^{kl})\partial_\mu\phi_{jkl}]
\epsilon.
\end{equation}
In the above, the lower dimensional quantities are related to their $D=11$
counterparts through
\begin{eqnarray}
&&\psi_\mu=\Delta^{\fft1{4(d-2)}}\left(\Psi^{(11)}_\mu+\fft1{d-2}\gamma_\mu
\Gamma^i\Psi^{(11)}_i\right),\qquad
\lambda_i=\Delta^{\fft1{4(d-2)}}\Psi^{(11)}_i,\nonumber\\
&&\epsilon=\Delta^{\fft1{4(d-2)}}\epsilon^{(11)},\nonumber\\
&&Q^{ab}_\mu=e^{i[a}\partial_\mu e_i{}^{b]},\qquad
P_{\mu\,ij}=e_{(i}^a\partial_\mu e_{j)\,a}.
\end{eqnarray}

This decomposition is suggestive of a
generalized structure group with connection given by ${\hat D}_\mu$.  
However one additional requirement is necessary before declaring this 
an enlargement of $\SO(d-1,1)\times \SO(11-d)$, and that is to ensure 
that the algebra generated by $\Gamma_{ab}$ and $\Gamma_{abc}$ closes 
within itself.  Along this line, we note that the commutators of these 
internal Dirac matrices have the schematic structure %
\begin{equation}
[\Gamma^{(2)},\Gamma^{(2)}]=\Gamma^{(2)},\qquad
[\Gamma^{(2)},\Gamma^{(3)}]=\Gamma^{(3)},\qquad
[\Gamma^{(3)},\Gamma^{(3)}]=\Gamma^{(6)}+\Gamma^{(2)}.
\label{eq:diralg}
\end{equation}
Here the notation $\Gamma^{(n)}$ indicates the antisymmetric product of
$n$ Dirac matrices, and the right hand sides of the commutators only
indicate what possible terms may show up.  The first commutator above
merely indicates that the $\Gamma_{ab}$ matrices provide a
representation of the Riemannian $\SO(11-d)$ structure group.

For $d\ge6$, the internal space is restricted to five or fewer
dimensions.  In this case, the antisymmetric product $\Gamma^{(6)}$
cannot show up, and the algebra clearly closes on $\Gamma^{(2)}$ and
$\Gamma^{(3)}$.  Working out the extended structure groups for these
cases results in the expected Cremmer and Julia groups listed in the
first four lines in the second column of Table \ref{gen}.  A similar analysis
follows for $d\le5$.  However, in this case, we must also dualize an
additional set of fields to see the hidden symmetries.  For $d=5$, an
additional scalar arises from the dual of 
$F_{\mu\nu\rho\sigma}$ ; this yields an addition to (\ref{hatD}) of the 
form ${\hat D}_\mu^{\rm additional}=\fft1{4!} 
\epsilon_\mu{}^{\nu\rho\sigma\lambda} 
F_{\nu\rho\sigma\lambda}\Gamma_{123456}$.  This $\Gamma^{(6)}$ term is 
precisely what is necessary for the closure of the algebra of 
(\ref{eq:diralg}).  Of course, in this case, we must also make note of 
the additional commutators %
\begin{equation}
[\Gamma^{(2)},\Gamma^{(6)}]=\Gamma^{(6)},\qquad
[\Gamma^{(3)},\Gamma^{(6)}]=\Gamma^{(7)}+\Gamma^{(3)},\qquad
[\Gamma^{(6)},\Gamma^{(6)}]=\Gamma^{(10)}+\Gamma^{(6)}+\Gamma^{(2)}.
\label{eq:gam6com}
\end{equation}
However neither $\Gamma^{(7)}$ nor $\Gamma^{(10)}$ may show up in $d=5$ for
dimensional reasons.

The analysis for $d=4$ is similar; however here
${\hat D}_\mu^{\rm 
additional}=\fft1{3!}\epsilon_\mu{}^{\nu\rho\sigma}e^{ia}F_{\nu\rho\sigma 
i} \Gamma_a\Gamma_{1234567}$.  Closure of the algebra on 
$\Gamma^{(2)}$, $\Gamma^{(3)}$ and $\Gamma^{(6)}$ then follows 
because, while $\Gamma^{(7)}$ may in principle arise in the middle 
commutator of (\ref{eq:gam6com}), it turns out to be kinematically 
forbidden.  For $d=3$, on the other hand, in additional to a 
contribution ${\hat D}_\mu^{\rm additional} 
=\fft1{2!\cdot2!}\epsilon_\mu{}^{\nu\rho}e^{ia}e^{jb}F_{\nu\rho ij} 
\Gamma_{ab}\Gamma_{12345678}$, one must also dualize the Kaluza-Klein 
vectors $g_\mu{}^i$.  Doing so gives rise to a $\Gamma^{(7)}$ in the 
generalized connection which, in addition to the previously identified 
terms, completes the internal structure group to $\SO(16)$.

The remaining three cases, namely $d=2$, $d=1$ and $d=0$ fall
somewhat outside the framework presented above.  This is because in
these low dimensions the generalized connections ${\hat D}_\mu$ derived via 
reduction are partially incomplete.  For $d=2$, we find %
\begin{equation}
{\hat D}_\mu^{(d=2)}=\partial_{\mu}+\omega_\mu{}^{\alpha\beta} 
\gamma_{\alpha\beta}+Q_\mu{}^{ab}\Gamma_{ab}+\ft19(\delta_\mu^\nu- 
\ft12\gamma_\mu{}^\nu)e^{ia}e^{jb}e^{kc}\partial_\nu\phi_{ijk}\Gamma_{abc},
\label{eq:d2omega}
\end{equation}
where $\gamma_{\mu\nu}=-\fft12\epsilon_{\mu\nu}
(\epsilon^{\alpha\beta}\gamma_{\alpha\beta})$ is necessarily proportional
to the two-dimensional chirality matrix.  Hence from a two-dimensional
point of view, the scalars from the metric enter non-chirally, while the
scalars from $F_{(4)}$ enter chirally.  Taken together, the generalized
connection (\ref{eq:d2omega}) takes values in $\SO(16)_+\times
\SO(16)_-$, which we regard as the enlarged structure group.  However not
all generators are present because of lack of chirality in
the term proportional to $Q_\mu{}^{ab}$.  Thus at this point the
generalized structure group deviates from the hidden symmetry group,
which would be an infinite dimensional subgroup of affine ${\rm E}_8$.
Similarly, for $d=1$, closure of the derivative ${\hat D}_\mu^{(d=1)}$ 
results in an enlarged $\SO(32)$ structure group.  However this is not 
obviously related to any actual hidden symmetry of the $1/10$ split.  
The $d=0$ case is subject to the same caveats as the $d=1$ and $d=2$ 
cases: not all group generators are present in the covariant 
derivative.  $\SL(32,\R)$ requires $\{\Gamma^{(1)},\Gamma^{(2)}, 
\Gamma^{(3)},\Gamma^{(4)},\Gamma^{(5)}\}$ whereas only 
$\{\Gamma^{(2)}, \Gamma^{(3)},\Gamma^{(5)}\}$ appear in the covariant 
derivative.

Next we consider a timelike reduction for which we simply interchange a
time and a space direction in the above analysis.  This results in an
internal Clifford algebra with signature $(10-d,1)$, and yields the
extended symmetry groups indicated in the fourth column of Table \ref{gen}.
The same caveats concerning $d=2,1,0$ apply in the timelike case.

Turning finally to the null case, we may replace one of the internal
Dirac matrices with $\Gamma_+$ (where $+$, $-$ denote light-cone
directions).  Since $(\Gamma_+)2=0$, this indicates that the extended
structure groups for the null case are contractions of the corresponding
spacelike (or timelike) groups.  In addition, by removing $\Gamma_+$
from the set of Dirac matrices, we essentially end up in the case of
one fewer compactified dimensions.  As a result, the $G(null)$ group
in $d$-dimensions must have a semi-direct product structure involving
the $G(spacelike)$ group in $(d+1)$-dimensions.  Of course, these
groups also contain the original $\ISO(10-d)$ structure group as a
subgroup.  The resulting generalized structure groups are given in the
third column of Table \ref{gen}.  Once again, the same caveats concerning
$d=2,1,0$ apply.

Spacelike reductions of D=11 supergravity may be found in the paper 
of Cremmer and Julia \cite{Cremmerjulia}, null reductions in the paper of 
Duff and Liu \cite{Duff:2003ec} and timelike reductions in the paper 
of Hull and Julia \cite{Hulljulia}.  Some of the 
noncompact groups appearing in the Table may be unfamiliar, but a nice 
discussion of their properties may be found in the book by Gilmore \cite{Gilmore}.

\subsection{The complete uncompactified D=11 theory}
\label{complete}

Following Cremmer and Julia's spacelike reduction, the question
was then posed: do these symmetries appear magically only after 
dimensional reduction, or were they already present in the full 
uncompactified and untruncated $D=11$ theory?  The question was 
answered by de Wit and Nicolai who made a $d/(11-d)$ split and 
fixed the gauge by setting to zero the off-diagonal components of the 
elfbein.  They showed that in the resulting field equations the local 
symmetries are indeed already present, but the global symmetries are 
not.  For example, after making the split $\SO(10,1) \supset \SO(3,1) 
\times \SO(7)$, we find the enlarged symmetry $\SO(3,1) \times 
\SU(8)$.  There is no global ${\rm E}_7$ invariance (although the 70 
internal components of the metric and 3-form may nevertheless be 
assigned to an ${\rm E}_7/\SU(8)$ coset).  Similar results were found 
for other values of $d$: in each case the internal subgroup 
$\SO(11-d)$ gets enlarged to some compact group $G(spacelike)$ while 
the spacetime subgroup $\SO(d-1,1)$ remains intact%
\footnote{We keep the terminology ``spacetime'' and ``internal''
even though no compactification or dimensional reduction is implied.}.
Here we ask instead whether there are hidden {\it spacetime} symmetries.
This is a question that could have been asked long ago, but we suspect
that people may have been inhibited by the Coleman-Mandula theorem
which forbids combining spacetime and internal symmetries.  
However, this is a statement about Poincar{e} symmetries of the 
S-matrix and here we are concerned with Lorentz symmetries of the 
equations of motion, so there will be no conflict.

The explicit demonstration of $G(spacelike)$ invariance by de Wit and
Nicolai is very involved, to say the least.  However, the result is
quite simple: one finds the same $G(spacelike)$ in the full
uncompactified $D=11$ theory as was already found in the spacelike
dimensional reduction of Cremmer and Julia.  Here we content ourselves
with the educated guess that the same logic applies to $G(timelike)$
and $G(null)$: they are the same as what one finds by timelike and
null reduction, respectively. The claim that the null and timelike
symmetries are present in the full theory and not merely in its
dimensional reductions might be proved by repeating the spacelike
calculations of de Wit and Nicolai with the appropriate change of
$\Gamma$ matrices.  So we propose
that, after making a $d/(11-d)$ split, the Lorentz subgroup
$G=\SO(d-1,1) \times \SO(11-d)$ can be enlarged to the
generalized structure groups ${\cal G}=\SO(d-1,1) \times
G(spacelike)$, ${\cal G}= \ISO(d-1) \times G(null)$ and ${\cal G}= \SO(d)
\times G(timelike)$.

\begin{table}[t]
\begin{center}
\begin{tabular}{c|ccc}
$d/(11-d)$&$G(spacelike)$&$G(null)$&$G(timelike)$\\
\hline
11/0&$\{1\}$&$\{1\}$&$\{1\}$\\
10/1&$\{1\}$&$\{1\}$&$\{1\}$\\
9/2&$\SO(2)$&$\R$&$\SO(1,1)$\\
8/3&$\SO(3) \times \SO(2)$&$\ISO(2) \times \R$&$\SO(2,1) \times \SO(1,1)$\\
7/4&$\SO(5)$&$[\SO(3) \times \SO(2)] \ltimes
\R6_{(3,2)}$&$ \SO(3,2)$\\
6/5&$\SO(5) \times \SO(5)$&$\SO(5) \ltimes \R^{10}_{(10)}$&$\SO(5,\C)$\\
5/6&${\rm USp}(8)$&$[\SO(5) \times \SO(5)] \ltimes
\R^{16}_{(4,4)}$&${\rm USp}(4,4)$\\
4/7&$\SU(8)$&${\rm USp}(8)\ltimes \R^{27}_{(27)}$&$\SU^*(8)$\\
3/8&$\SO(16)$&$[\SU(8) \times \U(1)]\ltimes
                        \R^{56}_{(28_{1/2},\overline{28}_{-1/2})}$&$\SO^*(16)$\\
\hline
2/9&$\SO(16) \times \SO(16)$&$\SO(16) \ltimes \R^{120}_{(120)}$&$\SO(16,\C)$\\
1/10&$\SO(32)$&$[\SO(16)\times\SO(16)]\ltimes\R^{256}_{(16,16)}$&$\SO(16,16)$\\
0/11&$\SL(32,\R)$&$\SL(32,\R)$&$\SL(32,\R)$
\end{tabular}
\end{center}
\caption{The generalized structure groups are given by ${\cal
G}=\SO(d-1,1) \times G(spacelike)$, ${\cal G}= \ISO(d-1) \times
G(null)$ and ${\cal G}= \SO(d) \times G(timelike)$.}
\label{gen}
\end{table}

As we have seen, for $d>2$ the groups $G(spacelike)$, $G(timelike)$ and
$G(null)$ are the same as those obtained from dimensional reductions.
For the purposes of this section, however, their physical interpretation
is very different.  They are here proposed as symmetries of the full
$D=11$ equations of motion; there is no compactification involved,
whether toroidal or otherwise.  (Note that by postulating that the generalized
structure groups survive as hidden symmetries of the full
uncompactified theory, we avoid the undesirable features associated
with compactifications including a timelike direction such as closed
timelike curves.)

For $d\leq 2$ it is less clear whether these generalized
structure groups are actually hidden symmetries.  Yet one might
imagine that there exists a yet-to-be-discovered
formulation of M-theory in which the $d=2$ and $d=1$ symmetries are
realized.  This would still be in keeping with the apparent need to 
make a non-covariant split and to make the corresponding gauge choice 
before the hidden symmetries emerge.  A yet bolder conjecture, due to 
Hull, 
requiring no non-covariant split or gauge choice since $d=0$ is that 
there exists a formulation of M-theory with the full 
$\SL(32,\R)$.  This proposal is nevertheless very attractive since 
$\SL(32,\R)$ contains all the groups in Table \ref{gen} as subgroups 
and would thus answer the question of whether all these symmetries are 
present at the same time.  This is an important issue deserving of 
further study.

We can apply similar logic to theories with fewer than 32
supersymmetries.  Of course, if M-theory really underlies all
supersymmetric theories then the corresponding vacua will all be
special cases of the above.  However, it is sometimes useful to focus
on such a sub-theory, for example the Type I and heterotic strings
with $N=16$.  Here $G(spacelike)= \SO(d) \times \SO(d)$, $G(null)=
\ISO(d-1) \times \ISO(d-1)$ and $G(timelike)=\SO(d-1,1) \times
\SO(d-1,1)$.

Finally, we emphasize that despite the $d/(11-d)$ split these symmetries
refer to the full equations of motion and not to any particular background
such as product manifolds.  This issue of specific solutions of these
equations is the subject of the next section.

Note that we have not considered the global symmetries such as $E_{7}$ 
for d=4, $E_{8}$ for d=3 and their infinite dimensional 
generalizations $E_{11-d}$ for $d\leq 2$. These appear after 
dimensional reduction but, according to de Wit 
and Nicolai, not even the finite dimensional examples are symmetries of 
the full uncompactified theory. Discrete subgroups, known as 
U-dualities, do appear in M-theory, but so far only as symmetries of 
toroidally compactified vacua, not as background-independent symmetries of the 
equations of motion. 
 
Hidden symmetries of the uncompactified $D=11$ equations, as opposed to 
their dimensional reduction, are discussed in the papers by Duff 
\cite{Fradkin}, de Wit and Nicolai \cite{Dewitnicolai2,Nicolai}, Duff and Liu 
\cite{Duff:2003ec}, Hull \cite{Hull:2003mf} and Keurentjes 
\cite{Keurentjes1,Keurentjes2}.

U-duality conjectures in membrane and M-theory may be found in the papers of 
Duff and Liu \cite{U1} and Hull and Townsend \cite{U2}. For a recent 
discussion of $E_{11}$ see the paper by West \cite{West2}.


\section{Counting supersymmetries of D=11 vacua}
\label{counting}

\subsection{Holonomy and supersymmetry}
\label{Riemannian}

The equations of M-theory display the maximum number of supersymmetries
$N=32$, and so $n$, the number of supersymmetries preserved by a
particular vacuum, must be some integer $0\leq n\leq 32$.  In vacua
with vanishing 4-form $F_{(4)}$, it is well known that $n$ is given by the
number of singlets appearing in the decomposition of the 32 of $\SO(1,10)$
under ${H} \subset \SO(1,10)$ where $H$ is the holonomy group of the
usual Riemannian connection (\ref{usual}).  This connection can account for vacua
with $n=0$, 1, 2, 3, 4, 6, 8, 16, 32.

Vacua with non-vanishing $F_{(4)}$ allow more exotic fractions of
supersymmetry, including $16<n<32$.  Here, however, it is necessary
to generalize the notion of holonomy to accommodate the generalized
connection (\ref{supercovariant}) that results from a non-vanishing $F_{(4)}$.  
As discussed by Duff and Liu, the number of M-theory vacuum 
supersymmetries is now given by the number of singlets appearing in 
the decomposition of the $32$ of ${\cal G}$ under $\Hol \subset \cal 
G$ where $\Hol$ is the generalized holonomy group and $\cal G$ is the 
generalized structure group.

In subsequent papers by Hull and by Papadopoulos and Tsimpis 
it was shown that $\cal G$ may be as large as $\SL(32,\R)$ and that 
an M-theory vacuum admits precisely $n$ Killing spinors iff 
\begin{equation}
\SL(31-n,\R) \ltimes (n+1)\R^{(31-n)} \supseteq\kern-12pt/\kern8pt {\Hol}
\subseteq  \SL(32-n,\R)\ltimes n\R^{(32-n)},
\label{n}
\end{equation}
{\it i.e.} the generalized holonomy is contained in $\SL(32-n,\R)\ltimes
n\R^{(32-n)}$ but is not contained in $\SL(31-n,\R) \ltimes
(n+1)\R^{(31-n)}$.

We recall that the number of supersymmetries preserved by an M-theory 
background depends on the number of covariantly constant spinors, %
\begin{equation}
{\cal D}_{M}\epsilon=0,
\label{integrability1}
\end{equation}
called {\it Killing} spinors.  It is the presence of the terms involving
the 4-form $F_{(4)}$ in (\ref{supercovariant}) that makes this counting 
difficult.  So let us first examine the simpler vacua for which 
$F_{(4)}$ vanishes.  Killing spinors then satisfy the integrability 
condition %
\begin{equation}
[{D}_{M}, {D}_{N}] \epsilon=\frac{1}{4}R_{MN}{}^{AB}\Gamma_{AB}\epsilon=0,
\label{integrability2}
\end{equation}
where $R_{MN}{}^{AB}$ is the Riemann tensor.  The subgroup of
$\Spin(10,1)$ generated by this linear combination of $\Spin(10,1)$ generators
$\Gamma_{AB}$ corresponds to the ${\it holonomy}$ group ${H}$ of the
connection $\omega_{M}$.  We note that the same information is
contained in the first order Killing spinor equation (\ref{integrability1})
and second-order integrability condition (\ref{integrability2}).  One
implies the other, at least locally.  The number of
supersymmetries, $n$, is then given by the number of singlets
appearing in the decomposition of the $32$ of $\Spin(10,1)$ under
${H}$.  In Euclidean signature, connections satisfying
(\ref{integrability2}) are automatically Ricci-flat and hence solve
the field equations when $F_{(4)}=0$.  In Lorentzian signature, however,
they need only be Ricci-null so Ricci-flatness has to be imposed as an
extra condition.  In Euclidean signature, the holonomy groups have
been classified.  In Lorentzian signature, much less is known but 
the question of which subgroups ${H}$ of $\Spin(10,1)$ leave a spinor 
invariant has been answered by Bryant.  There are two sequences 
according as the Killing vector 
$v_{A}=\overline{\epsilon}\,\Gamma_{A}\epsilon$ is timelike or null.  
Since $v^{2} \leq 0$, the spacelike $v_A$ case does not arise.  The 
timelike $v_A$ case corresponds to static vacua, where ${H} \subset 
\Spin(10) \subset \Spin(10,1)$ while the null case to non-static vacua 
where ${H} \subset \ISO(9) \subset \Spin(10,1)$.  It is then possible 
to determine the possible $n$-values and one finds $n=2$, 4, 6, 8, 16, 
32 for static vacua, and $n=1$ 2, 3, 4, 8, 16, 32 for non-static vacua 
as shown in 
Table~\ref{static}, and $n=1$, 2, 3, 4, 8, 16, 32 for non-static 
vacua, as shown in Table~\ref{wave}.

\begin{table}[t]
\begin{center}
\begin{tabular}{c|cc}
$d/(11-d)$&$H \subset \SO(11-d)\subset\Spin(10)$& $n$\\
\hline
7/4&$\SU(2) \cong \Sp(2)$& $16$\\
5/6&$\SU(3)$& $8$\\
4/7&${\rm G}_2$& $4$\\
3/8&$\SU(2) \times \SU(2)$& $8$\\
&$\Sp(4)$& $6$\\
&$\SU(4)$& $4$\\
&$\Spin(7)$& $2$\\
1/10&$\SU(2) \times \SU(3)$& $4$\\
&$\SU(5)$& $2$
\end{tabular}
\end{center}
\caption{Holonomy of static M-theory vacua with $F_{(4)}=0$ and their
supersymmetries.}
\label{static}
\end{table}

\begin{table}[t]
\begin{center}
\begin{tabular}{c|cc}
$d/(11-d)$&$H \subset \ISO(d-1)\times\ISO(10-d)\subset\Spin(10,1)$& $n$\\
\hline
10/1&$\R^{9}$ & $16$\\
6/5&$\R^{5}\times(\SU(2) \ltimes \R^{4})$ & $8$\\
4/7&$\R^{3}\times(\SU(3) \ltimes \R^{6})$ & $4$\\
3/8&$\R^{2}\times({\rm G}_2 \ltimes \R^{7})$ & $2$\\
2/9&$\R\times(\SU(2) \ltimes \R^{4}) \times (\SU(2) \ltimes \R^{4})$ & $4$\\
&$\R\times(\Sp(4) \ltimes \R^{8})$ & $3$\\
&$\R\times(\SU(4) \ltimes \R^{8})$ & $2$\\
&$\R\times(\Spin(7) \ltimes \R^{8})$ & $1$
\end{tabular}
\end{center}
\caption{Holonomy of non-static M-theory vacua with $F_{(4)}=0$ and their
supersymmetries.}
\label{wave}
\end{table}

The allowed $n$ values for Riemannian connections may be found in the 
papers of Acharya et al \cite{Acharya:1998yv,Acharya:1998st} and by 
Figueroa-O'Farrill \cite{Fig}.

\subsection{Generalized holonomy}
\label{Generalized}

In general we want to include vacua with $F_{(4)}\neq 0$.  Such vacua
are physically interesting for a variety of reasons. In particular, they
typically have fewer moduli than their zero $F_{(4)}$ counterparts.  
Now, however, we face the problem that the connection in 
(\ref{supercovariant}) is no longer the spin connection to which the 
bulk of the mathematical literature on holonomy groups is devoted.  In 
addition to the $\Spin (10,1)$ generators $\Gamma_{AB}$, it is 
apparent from (\ref{supercovariant}) that there are terms involving 
$\Gamma_{ABC}$ and $\Gamma_{ABCDE}$.  In fact, the generalized 
connection takes its values in $\SL(32,\R)$.  Note, however, that some 
generators are missing from the covariant derivative.  Denoting the 
antisymmetric product of $k$ Dirac matrices by $\Gamma^{(k)}$, the 
complete set of $\SL(32,\R)$ generators include 
$\{\Gamma^{(1)},\Gamma^{(2)}, 
\Gamma^{(3)},\Gamma^{(4)},\Gamma^{(5)}\}$ whereas only 
$\{\Gamma^{(2)}, \Gamma^{(3)},\Gamma^{(5)}\}$ appear in the covariant 
derivative.  Another way in which generalized holonomy differs from 
the Riemannian case is that, although the vanishing of the covariant 
derivative of the spinor implies the vanishing of the commutator, the 
converse is not true, as discussed below.

This generalized connection can preserve exotic fractions of
supersymmetry forbidden by the Riemannian connection.  For example,
M-branes at angles include $n$=5, 11-dimensional pp-waves include 
$n=18$, 20, 22, 24, 26, squashed $N(1,1)$ spaces and M5-branes in 
a pp-wave background include $n=12$ and G{o}del universes 
include $n=14$, 18, 20, 22, 24.  However, we can attempt to 
quantify this in terms of generalized holonomy groups
\footnote{In these lectures we focus on $D=11$ but similar generalized holonomy
can be invoked to count $n$ in Type IIB vacua, which include 
pp-waves with $n=28$.}.  %

Generalized holonomy means that one can assign a holonomy ${\cal H}
\subset {\cal G}$ to the generalized connection appearing in the
supercovariant derivative ${\cal D}$ where ${\cal G}$ is the generalized
structure group.  The number of unbroken supersymmetries is then given by
the number of ${\cal H}$ singlets appearing in the decomposition of the
32 dimensional representation of ${\cal G}$ under $\Hol \subset {\cal G}$.

For generic backgrounds we require that ${\cal G}$ be the full $\SL(32,\R)$
while for special backgrounds smaller ${\cal G}$ are sufficient.  
To see this, let us write the supercovariant derivative as %
\begin{equation}
{\cal D}_{M}=\hat D_{M}+X_{M},
\label{split2}
\end{equation}
for some other connection $\hat D_{M}$ and some covariant $32 \times 32$
matrix $X_{M}$. If we now specialize to backgrounds satisfying
\begin{equation}
X_{M}\epsilon=0,
\label{X}
\end{equation}
then the relevant structure group is ${\hat G} \subseteq {\cal G}$.

Consider, for example, the connection ${\hat D}$ arising in dimensional
reduction of $D=11$ supergravity (\ref{hatD}).  The condition (\ref{X}) is 
just $\delta \lambda_{i}=0$ where $\lambda_{i}$ are the dilatinos of 
the dimensionally reduced theory.  In this case, the generalized 
holonomy is given by ${\hat H} \subseteq {\hat G}$ where the various 
$\hat G$ arising in spacelike, null and timelike compactifications are 
tabulated in Table \ref{gen} for different numbers of the compactified 
dimensions.

Another way in which generalized holonomy differs from Riemannian
holonomy is that, although the vanishing of the covariant derivative
implies the vanishing of the commutator, the converse is not true.
Consequently, the second order integrability condition alone may be a
misleading guide to the generalized holonomy group $\cal H$.

To illustrate this, we consider Freund-Rubin vacua with $F_{(4)}$ 
given by %
\begin{equation}
F_{\mu\nu\rho\sigma}=3m\epsilon_{\mu\nu\rho\sigma},
\end{equation}
where $\mu=0,1,2,3$ and $m$ is a constant with the dimensions of mass.
This leads to an $AdS_{4} \times X^{7}$ geometry.  For such a product
manifold , the supercovariant derivative splits as
\begin{equation}
{\cal D}_{\mu}= D_{\mu}+m\gamma_{\mu}\gamma_{5}
\end{equation}
and
\begin{equation}
{\cal D}_{m}= D_{m}-\ft{1}{2}m\Gamma_{m},
\label{covariant1}
\end{equation}
and the Killing spinor equations reduce to
\begin{equation}
{\cal D}_{\mu}\epsilon(x) = 0
\end{equation}
and
\begin{equation}
{\cal D}_{m}\eta(y)= 0.
\label{Killing}
\end{equation}
Here $\epsilon(x)$ is a 4-component spinor and $\eta(y)$ is an
8-component spinor, transforming with Dirac matrices $\gamma_\mu$ and
$\Gamma_m$ respectively.  The first equation is satisfied automatically
with our choice of $AdS_{4}$ spacetime and hence the number of $D=4$
supersymmetries, $0\leq N \leq 8$, devolves upon the number of
Killing spinors on $X^{7}$.  They satisfy the integrability condition
\begin{equation}
[{\cal D}_{m}, {\cal D}_{n}] \eta=
-\frac{1}{4}C_{mn}{}^{ab}\Gamma_{ab}\eta=0,
\label{integrability}
\end{equation}
where $C_{mn}{}^{ab}$ is the Weyl tensor. Owing to this generalized
connection, vacua with $m\neq 0$ present subtleties and novelties not
present in the $m=0$ case, for example the phenomenon of
{\it skew-whiffing}.  For each Freund-Rubin compactification, one may obtain another by
reversing the orientation of $X^{7}$.  The two may be distinguished by
the labels {\it left} and {\it right}.  An equivalent way to obtain
such vacua is to keep the orientation fixed but to make the
replacement $m\rightarrow -m$ thus reversing the sign of $F_{4}$.  So
the covariant derivative (\ref{covariant1}), and hence the condition
for a Killing spinor, changes but the integrability condition
(\ref{integrability}) remains the same.  With the exception of the
round $S^{7}$, where both orientations give $N=8$, at most one
orientation can have $N \geq 0$.  This is the {\it skew-whiffing
theorem}. 

The squashed $S^{7}$ provides a non-trivial example
: the left squashed
$S^{7}$ has $N=1$ but the right squashed $S^{7}$ has $N=0$.  Other
examples are provided by the left squashed $N(1,1)$ spaces, one of 
which has $N=3$ and the other $N=1$, while the right squashed 
counterparts both have $N=0$.  (Note, incidentally, that $N=3$ {\it 
i.e.} $n=12$ can never arise in the Riemannian case.)

All this presents a dilemma.  If the Killing spinor
condition changes but the integrability condition does not, how does
one give a holonomic interpretation to the different supersymmetries?
We note that in (\ref{covariant1}), the $SO(7)$ generators $\Gamma_{ab}$,
augmented by the
presence of $\Gamma_{a}$, together close on $SO(8)$ .
Hence the generalized holonomy group satisfies ${\cal H}\subset SO(8)$.
We now ask how the $8$ of $SO(8)$ decomposes under ${\cal H}$.  In the
case of the left squashed $S^{7}$, ${\cal H}= SO(7)^{-}$, ${ 8
\rightarrow 1+7}$ and $N=1$, but for the right squashed $S^{7}$, ${\cal
H}= SO(7)^{+}$, ${ 8 \rightarrow 8}$ and $N=0$.  From the integrability
condition alone, however, we would have concluded naively that ${\cal
H}=G_{2}$ and that both orientations give $N=1$.

Another context in which generalized holonomy may prove important is
that of higher loop corrections to the M-theory Killing spinor equations with
or without the presence of non-vanishing $F_{(4)}$.  Higher loops 
yield non-Riemannian corrections to the supercovariant derivative, 
even for vacua for which $F_{(4)}=0$, thus rendering the Berger 
classification inapplicable.  Although the Killing spinor equation 
receives higher order corrections, so does the metric, ensuring, for 
example, that $H=G_{2}$ Riemannian holonomy 7-manifolds still yield 
$N=1$ in $D=4$ when the non-Riemannian corrections are taken into 
account.  This would require a generalized holonomy ${\cal H}$ for 
which the decomposition $8 \rightarrow 1+7$ continues to hold.

 Generalized holonomy is discussed in the 
papers of Duff and Stelle \cite{Duffstelle}, Duff \cite{Duff}, Duff and Liu 
\cite{Duff:2003ec}, Hull \cite{Hull:2003mf}, Papadopoulos and Tsimpis 
\cite{Papadopoulos:2003pf,Papadopoulos:2003jk}, Batrachenko, Duff, Liu 
and Wen \cite{Batrachenko:2003ng}, Bandos, de Azc{a}rraga, Izquierdo, 
Lukierski, 
Pic{o}n and Varela \cite{Bandos,Bandos:2001pu} and Keurentjes \cite{Keurentjes1,Keurentjes2}.

Skew-whiffing is discussed in the paper and Physics Report by Duff, 
Nilsson and Pope \cite{Duffnilssonpopesuper,Duffnilssonpopekaluza} and 
the paper of van Nieuwenhuizen and Warner \cite{vanNwarner}.
The squashed $S^{7}$ may be found in the papers of Awada, Duff and Pope 
\cite{Awadaduffpope} and Duff, Nilsson and Pope 
\cite{Duffnilssonpopesuper}.  For the result that $SO(7)$ generators $\Gamma_{ab}$, 
augmented by presence of $\Gamma_{a}$, together close on $SO(8)$ 
see the paper by Castellani, D'Auria, Fre and van 
Nieuwenhuizen \cite{Castellani1}.

Higher loop corrections to the Killing spinor equation are treated in 
the paper by Lu, Pope, Stelle and Townsend \cite{Lu:2003ze}.

\subsection{Specific examples}

In Table \ref{tab:1} we tabulate the results of computations of this 
generalized holonomy for the $n=16$ examples of the M2-brane, the 
M5-brane, the M-wave (MW) and the M-monopole (MK), and for a variety 
of their $n=8$ intersections: M5/MK, M2/MK/MK, M2/MK, M2/MW, 
M5/MW,MW/MK and M2/M5.  As we can see, the generalized holonomy of M-theory solutions takes 
on a 
variety of guises. We make note of two features exhibited by these solutions.  Firstly, 
it is clear that many generalized holonomy groups give rise to the 
same number $n$ of supersymmetries.  This is a consequence of the fact 
that while $\Hol$ must satisfy the condition (\ref{n}), there are 
nevertheless many possible subgroups of $\SL(32-n,\R)\ltimes 
n\R^{(32-n)}$ allowed by generalized holonomy.  Secondly, as 
demonstrated by the plane wave solutions, knowledge of $\Hol$ by 
itself is insufficient for determining $n$; here $\Hol=\R^9$, while 
$n$ may be any even integer between 16 and 26.

\begin{table}[t]
\begin{tabular}{lll}
$n$&Background&Generalized holonomy\\
\hline
32&$\E^{1,10}$, AdS$_7\times S^4$, AdS$_4\times S^7$, Hpp&$\{1\}$\\
\hline
18,\ldots,26&plane waves&$\R^9$\\
\hline
16&M5&$\SO(5)\ltimes6\R^{4(4)}$\\
16&M2&$\SO(8)\ltimes12\R^{2(8_s)}$\\
16&MW&$\R^9$\\
16&MK&$\SU(2)$\\
\hline
8&M5/MK&$[\SO(5)\times\SU(2)]\ltimes6\R^{2(4,1)+(4,2)}$\\
8&M2/MK/MK&$[\SO(8)\times\SU(2)\times\SU(2)\ltimes3\R^{(8_s,2,2)}]
\ltimes6\R^{2(8_s,1,1)}$\\
8&M2/MK&$[\SO(8)\times\SU(2)\ltimes3\R^{2(8_s,2)}]\ltimes6\R^{2(8_s,1,1)}$\\
8&M2/MW&$[\SO(8)\times\SL(16,\R)\ltimes \R^{(8,16)}]\ltimes8\R^{(8,1)
+(1,16)}$\\
8&M5/MW&$[\SO(5)\times\SU^*(8)\ltimes4\R^{(4,8)}]\ltimes8\R^{2(4,1)+2(1,8)}$\\
8&MW/MK&$\R^5\times(\SU(2)\ltimes\R^{2(2)})$\\
8&M2/M5&$\SL(24,\R)\ltimes8\R^{24}$\\
\hline
\end{tabular}
\caption{Generalized holonomies of the objects investigated in the text.
For $n=16$, we have $\Hol\subseteq\SL(16,\R)\ltimes16\R^{16}$, while
for $n=8$, it is instead $\Hol\subseteq\SL(24,\R)\ltimes8\R^{24}$.}
\label{tab:1}
\end{table}

What this indicates is that, at least for counting supersymmetries, it
is important to understand the embedding of $\Hol$ in $\cal G$.  In
contrast to the Riemannian case, different embeddings of $\Hol$ yield
different possible values of $n$.  Although this appears to pose a
difficulty in applying the concept of generalized holonomy towards
classifying supergravity solutions, it may be possible that a better
understanding of the representations of non-compact groups will nevertheless
allow progress to be achieved in this direction.

While the full generalized holonomy involves several factors, the transverse
(or $\hat D$) holonomy is often simpler, {\it e.g.} $SO(5)$ for the M5 and
$SO(8)$ for the M2.  The results summarized in table~\ref{tab:1} are
suggestive that the maximal compact subgroup of $\Hol$, which must be
contained in $\SL(32-n,\R)$, is often sufficient to determine the number of
surviving supersymmetries.  For example, the M2/MK/MK solution may be
regarded as a 3/8 split, with a hyper-Kahler eight-dimensional
transverse space.  In this case, the $\hat D$ structure group is $\SO(16)$,
and the 32-component spinor decomposes under $\SO(32)\supset
\SO(16)\supset\SO(8)\times\SU(2)\times\SU(2)$ as $32\to2(16)\to
2(8,1,1)+2(1,2,2)+8(1,1,1)$ yielding eight singlets.  Similarly, for the
M5/MW intersection, we consider a 2/9 split, with the wave running along
the two-dimensional longitudinal space.  Since the $\hat D$ structure
group is $\SO(16)\times \SO(16)$ and the maximal compact subgroup of
$\SU^*(8)$ is ${\rm USp}(8)$, we obtain the decomposition
$32\to(16,1)+(1,16)\to4(4,1)+(1,8)+8(1,1)$ under $\SO(32)\supset
\SO(16)\times\SO(16)\supset\SO(5)\times{\rm USp}(8)$.  This again yields
$n=8$.  Note, however, that this analysis fails for the plane waves, as
$\R^9$ has no compact subgroups.

Ultimately, one would hope to achieve a complete classification of M-theory
vacua, either through generalized holonomy or other means.  In this regard,
one must also include the effects of higher order corrections and perhaps
additional contributions beyond the supergravity itself.

\subsection{The full M(onty)?}
\label{M}

In sections \ref{hidden} and \ref{counting} we have focused on the low 
energy limit of M-theory, but since the reasoning is based mainly on group 
theory, it seems reasonable to promote it to the full M-theory.
Similar reasoning can be applied to M-theory in signatures
(9,2) and (6,5), the so-called M$^\prime$ and M$^*$ theories, but the groups 
will be different.
When counting the $n$ value of a particular vacuum, however, we should be
careful to note the phenomenon of {\it supersymmetry without
supersymmetry}, where the supergravity approximation may fail to
capture the full supersymmetry of an M-theory vacuum. For example,
vacua related by T-duality and S-duality must, by definition, have the
same $n$ values. Yet they can appear to be different in
supergravity if one fails to take into account winding
modes and non-perturbative solitons. So more work is needed to verify
that the $n$ values found so far in $D=11$ supergravity exhaust those of
M-theory.

A different approach to supersymmetric vacua in M-theory is 
through the technique of $G$-structures. Hull has suggested that $G$-structures may be better 
suited to finding supersymmetric solutions whereas generalized holonomy may be better 
suited to classifying them. In any event, it would be useful to 
establish a dictionary for translating one technique into the other. 

Ultimately, one would hope to achieve a complete classification of
vacua for the full M-theory.  
In this regard, one must at least include the effects of M-theoretic corrections to the 
supergravity field equations and Killing spinor equations and perhaps 
even go beyond the geometric picture altogether. It seems likely, 
however, that counting supersymmetries by the number of singlets 
appearing in the decomposition $32$ of $\SL(32,\R)$ under 
$\Hol \subset \SL(32,\R)$ will continue to be valid.

The various spacetime signatures in which M-theory can be formulated 
is discussed in the paper by Blencowe and Duff 
\cite{Blencowe:1988sk}.  M$^\prime$ and M$^*$ theories are treated 
in \cite{Hull:1998ym}. Supersymmetry without supersymmetry may be 
found in the papers of Duff, Lu and Pope \cite{DLP1,DLP2}.
For $G$-structures, see the papers by Gauntlett, Martelli, Pakis, Sparks 
and Waldram 
\cite{Gauntlett:2002fz,Gauntlett:2003wb,Gauntlett:2004zh,Gauntlett:2002sc,Martelli:2003ki} 
and by Hull \cite{Hull:2003mf}. 
Connections between generalized holonomy and G-structures in theories 
with 8 supercharges are discussed in the paper by Batrachenko and Wen 
\cite{Batrachenko}.


\section*{Acknowledgements}

The first lecture has benefitted from useful conversations with Gordy Kane and 
correspondence wuth Kelly Stelle.  The second lecture is based on work written 
with my collaborators Alex Batrachenko, Jim Liu and Steve Wen. Thanks 
to Arthur Greenspoon, who suggested several improvements to the manuscript. I 
am grateful to 
organizers of the school, Gerard `t Hooft and Nino Zichichi, for their kind 
hospitality in Erice. 



\newpage
\begin{thebibliography}{00}
    
\bibitem{Volkov1}    
D.V. Volkov and V.A. Soroka, 
{\sl Higgs effect for Goldstone particles with spin 1/2}, 
JETP Lett. 18 (1973) 312.

\bibitem{Volkov2}
D.V. Volkov and V.A. Soroka, 
{\sl Gauge fields for symmetry group with spinor parameters}, 
Theor. Math. Phys. 20 (1974) 829.

\bibitem{Ferrara}
S. Ferrara, D. Z. Freedman and P. van Nieuwenhuizen,
{\sl Progress toward a theory of supergravity},
Phys.  Rev.  {\bf D13} (1976) 3214.

\bibitem{Deser}
S. Deser and B. Zumino,
{\sl Consistent supergravity},
Phys.  Lett.  {\bf B62} (1976) 335.

\bibitem{Wess}
J. Wess and J. Bagger,
{\sl Supersymmetry and supergravity},
Princeton University Press (1983).

\bibitem{Gates}
S. J. Gates, Jr., M. T. Grisaru, M. Rocek and W. Siegel,
{\sl Superspace {\it or} one thousand and one lessons in
supersymmetry},
Benjamin Cummings (1983).

\bibitem{Srivastava}
Prem P. Srivastava,
{\sl Supersymmetry, superfields and supergravity},
Adam Hilger (1986).

\bibitem{West}
P. West,
{\sl Introduction to supersymmetry and supergravity},
World Scientific (1986).

\bibitem{Freund}
P. G. O. Freund,
{\sl Introduction to supersymmetry}
C. U. P. (1988).

\bibitem{Bailin}
D. Bailin and A. Love,
{\sl Supersymmetric gauge field theory and string theory},
I. O. P. (1994).
 
\bibitem{Weinberg}
S.  Weinberg,
{\sl The Quantum Theory of Fields, Volume III, Supersymmetry},
C.  U.  P. (2000).

\bibitem{soh}
M. Sohnius, {\sl Introducing Supersymmetry},
Phys.\ Rept.\  {\bf 128}, 39 (1985).

\bibitem{vanN}
P. Van Nieuwenhuizen,
{\sl Supergravity},
Phys. Rept. {\bf 68} (1981) 189.

\bibitem{Fayet}
P.~Fayet and S.~Ferrara, {\sl Supersymmetry},
Phys.\ Rept.\  {\bf 32}, 249 (1977).


\bibitem{Lykken}
J.~D.~Lykken,
{\sl Introduction to supersymmetry},
arXiv:hep-th/9612114.

\bibitem{Nanopoulos}
D. Nanopoulos,
{\sl Supersymmetric GUTS}
Phys.\ Rept.\  {\bf 105}, 71 (1984).

\bibitem{Nilles}
H. Nilles,
{\sl Supersymmetry, supergravity and particle physics},
Phys.\ Rept.\  {\bf 110}, 1 (1984).

\bibitem{Haber}
H.~E.~Haber and G.~L.~Kane,
{\sl The Search For Supersymmetry: Probing Physics Beyond The Standard Model},
Phys.\ Rept.\  {\bf 117}, 75 (1985).

\bibitem{Ellis}
J. Ellis,
{\sl Supersymmetry and grand unification},
hep-th/9512335.

\bibitem{Chung}
D. J. H. Chung, L. L. Everett, G. L. Kane, S. F. King, J. Lykken
and Lian-Tao Wang.  {\sl The soft supersymmetry-breaking lagrangian:
theory and applications}, MCTP-03-39.

\bibitem{Dine}
M. Dine,
{\sl TASI lectures on M Theory phenomenology},
hep-th/0003175. 

\bibitem{Ross}
G.~G.~Ross,
{\sl Supersymmetric models},
{\it Prepared for Les Houches Summer School in Theoretical Physics, 
Session 68: Probing the Standard Model of Particle Interactions, Les Houches, 
France, 28 Jul - 5 Sep 1997}
Ê

\bibitem{Raby}
S. Raby,
{\sl Desperately seeking supersymmetry},
hep-ph/0401169.

\bibitem{ultra}
M.~J.~Duff,
{\sl Ultraviolet divergences in extended supergravity},
CERN-TH-3232,
{\it Lectures given at Spring School on Supergravity, Trieste, Italy, Apr 22 - May 6, 1981}.

\bibitem{Lindstrom}
P.~S.~Howe and U.~Lindstrom,
{\sl Higher order invariants in extended supergravity},
Nucl.\ Phys.\ B {\bf 181}(1981) 487.

\bibitem{Bern}
Z.~Bern {\it et al.},
{\sl Counterterms in supergravity},
hep-th/0012230.

\bibitem{Howestelle} 
P.S. Howe and K.S. Stelle,
{\sl Supersymmetry counterterms revisited},
Phys. Lett. {\bf B554} (2003) 190.

\bibitem{Wittenfermion}
E. Witten,
{\sl Fermion quantum numbers in Kaluza-Klein theory},
in Proceedings of the Shelter Island II conference (1983) 227, eds
Khuri,
Jackiw, Weinberg and Witten (M. I. T. Press, 1985).

\bibitem{Duffnilssonpopekaluza}
 M. J. Duff, B. E. W. Nilsson and C. N. Pope,
{\sl Kaluza-Klein supergravity},
Phys. Rep. 130 (1986) 1.

\bibitem{Appelquist}
 T.  Appelquist, A.  Chodos and P.  G.  O.  Freund, 
{\sl Modern Kaluza-Klein theories},
(Addison-Wesley, 1987).

\bibitem{Castellani}
L. Castellani, R. D'Auria and P. Fre,
{\sl Supergravity and superstrings: a geometric perspective},
(in 3 volumes, World Scientific 1991).

\bibitem{Salamsezgin}
Abdus Salam and Ergin Sezgin,
{\sl Supergravities in diverse dimensions},
(World Scientific, 1989).

\bibitem{Klein1}
 M. J. Duff,
\newblock {\sl Kaluza-Klein theory in perspective},
\newblock in the Proceedings of the Nobel Symposium {\it Oskar Klein
Centenary}, Stockholm, September 1994, (Ed Lindstrom, World 
Scientific,
1995), hep-th/9410046.

\bibitem{Klein2}
 M. J. Duff,
\newblock {\sl The world in eleven dimensions: a tribute to Oskar Klein}, 
hep-th/0111237.

\bibitem{GSW}
M. B. Green, J. H. Schwarz and E. Witten,
\newblock {\sl Superstring theory},
\newblock Cambridge University Press (1987).

\bibitem{Polchinskistrings}
J. Polchinski,
\newblock {\sl String Theory}, Volumes I and II, Cambridge University 
Press, (1998).


\bibitem{Khuristring}
M.~J. Duff, R.~R. Khuri and J.~X. Lu,
\newblock {\sl String solitons},
\newblock Phys.  Rep.  {\bf 259} (1995) 213.

\bibitem{Schwarzpower}
J. H. Schwarz,
\newblock {\sl The power of $M$-theory},
\newblock Phys. Lett. {\bf B360} (1995) 13.

\bibitem{M}
M. J. Duff,
{\sl $M$-theory: the theory formerly known as strings},
Int. J. Mod. Phys. A11 (1996) 5623-5642, hep-th/9608117.

\bibitem{Duffworld1}
M. J. Duff,
\newblock {\sl The World in Eleven Dimensions: Supergravity,
Supermembranes and M-theory},
\newblock Institute of Physics Publishing (1999).

\bibitem{TownsendM}
P. K. Townsend,
\newblock {\sl Four lectures on $M$-theory},
\newblock hep-th/9612121.

\bibitem{Kaku1}
M. Kaku,
{\sl Introduction to superstrings and M-theory},
(Springer, 1999).

\bibitem{Kaku2}
M. Kaku,
{\sl Strings, conformal fields and M-theory},
(Springer, 1999).

\bibitem{Fifteen5}
M. J. Duff,
{\sl Supermembranes:  The first fifteen weeks},
 Class. Quant. Grav. {\bf 5}, 189 (1988).

\bibitem{Classical5}
M. J. Duff,
{\sl Classical and quantum supermembranes},
Class. Quant. Grav. {\bf 6}, 1577 (1989).

\bibitem{Duffsupermembranes5}
M. J. Duff,
\newblock {\sl Supermembranes},
\newblock Lectures given at the Theoretical Advanced Study Institute
in
Elementary Particle Physics (TASI 96), Boulder, 1996, hep-th/9611203.

\bibitem{Stelle5}
K. S. Stelle,
{\sl An introduction to $p$-branes},
In *Seoul/Sokcho 1997, Dualities in gauge and string theories 39-104.

\bibitem{Dbranes}
C.  Johnson,
{\sl D-branes},
C. U. P. 2003.

\bibitem{Ortin}
T. Ortin,
{\sl Gravity and strings},
C. U. P. 2004. 

\bibitem{Strathdee}
J. Strathdee,
{\sl Extended Poincar{e} supersymmetry},
Int. J. Mod. Phys. {\bf A2} (1987) 273.

\bibitem{Savoy}
S. Ferrara and C. A. Savoy, 
{\sl Representations of extended supersymmetry on one and two particle states}, 
in Supergravity 81, (Eds. Ferrara and Taylor, C. U. P,1982). 

\bibitem{Gauntletthull1}
J. P. Gauntlett and C. M. Hull,
{\sl BPS states with extra supersymmetry},
JHEP {\bf 0001}, 004 (2000) [hep-th/9909098].

\bibitem{Townsend}
P. K. Townsend,
{\sl M-theory from its superalgebra},
hep-th/9712004.

\bibitem{Duff}
M. J. Duff,
{\sl M-theory on manifolds of $G_{2}$ holonomy: the first twenty years},
hep-th/0201062.

\bibitem{Duff:2003ec}
M.~J.~Duff and J.~T.~Liu,
{\sl Hidden spacetime symmetries and generalized holonomy in M-theory},
hep-th/0303140.

\bibitem{Hull:2003mf}
C.~Hull,
{\sl Holonomy and symmetry in M-theory},
hep-th/0305039.

\bibitem{Papadopoulos:2003pf}
G.~Papadopoulos and D.~Tsimpis,
{\sl The holonomy of the supercovariant connection and Killing spinors},
JHEP {\bf 0307}, 018 (2003) [hep-th/0306117].

\bibitem{Papadopoulos:2003jk}
G.~Papadopoulos and D.~Tsimpis,
{\sl The holonomy of IIB supercovariant connection},
hep-th/0307127.

\bibitem{Batrachenko:2003ng}
A.~Batrachenko, M.~J.~Duff, J.~T.~Liu and W.~Y.~Wen, {\sl Generalized 
holonomy of M-theory vacua}, hep-th/0312165.

\bibitem{Bandos}
I.~A.~Bandos, J.~A.~de Azc{a}rraga, J.~M.~Izquierdo, M.~Pic{o}n and O.~Varela,
{\sl On BPS preons, generalized holonomies and D = 11 supergravities},
hep-th/0312266.

\bibitem{Keurentjes1}
A.~Keurentjes,
{\sl The topology of U-duality (sub-)groups},
hep-th/0309106. 
 
\bibitem{Keurentjes2}
A.~Keurentjes,
{\sl U-duality (sub-)groups and their topology},
hep-th/0312134.  

\bibitem{Wittensearch}
E. Witten,
{\sl Search for a realistic Kaluza-Klein theory},
Nucl.  Phys.  {\bf B186} (1981) 412-428.

\bibitem{Cremmerjuliascherk}
E. Cremmer, B. Julia and J. Scherk,
{\sl Supergravity theory in eleven-dimensions},
Phys.  Lett.  {\bf B76} (1978) 409.

\bibitem{Cremmerjulia}
E. Cremmer and B. Julia,
{\sl The $SO(8)$ supergravity},
Nucl.  Phys.  {\bf B159} (1979) 141.

\bibitem{Hulljulia}
C. Hull and B. Julia,
{\sl Duality and moduli spaces for time-like reductions},
Nucl.\ Phys.\ B {\bf 534}, 250 (1998) [hep-th/9803239].

\bibitem{Gilmore}
R. Gilmore,
{\it Lie groups, Lie algebras and some of their applications}
(Wiley, New York, 1974).

\bibitem{Fradkin}
M. J. Duff,
{\sl $E_8 \times SO(16)$ Symmetry of $d = 11$ Supergravity?},
in {\it Quantum Field Theory and Quantum Statistics, Essays in Honor of E.
S. Fradkin}, Batalin, Isham and Vilkovisky, eds. (Adam Hilger, 1986).

\bibitem{Dewitnicolai2}
B. de Wit and H. Nicolai,
{\sl $D=11$ supergravity with local $SU(8)$ invariance},
Nucl.\ Phys.\ B {\bf 274}, 363 (1986).

\bibitem{Nicolai}
H. Nicolai,
{\sl $D=11$ supergravity with local $SO(16)$ invariance},
Phys.\ Lett.\ B {\bf 187}, 316 (1987).

\bibitem{Fig}
J. Figueroa-O'Farrill,
{\sl Breaking the M-waves},
Class.\ Quant.\ Grav.\ {\bf 17}, 2925 (2000) [hep-th/9904124].

\bibitem{Berger}
M. Berger,
{\sl Sur les groupes d'holonomie homogene des variet{e}s a connexion
affine et des variet{e}s riemanniennes},
Bull. Soc. Math. France {\bf 83} (1955) 225.

\bibitem{Bryant}
R. L. Bryant,
{\sl Pseudo-Riemannian metrics with parallel spinor fields and
non-vanishing Ricci tensor},
math.DG/0004073.

\bibitem{Acharya:1998yv}
B.~S.~Acharya, J.~M.~Figueroa-O'Farrill and B.~Spence,
{\sl Planes, branes and automorphisms.  I: Static branes},
JHEP {\bf 9807}, 004 (1998) [hep-th/9805073].

\bibitem{Acharya:1998st}
B.~S.~Acharya, J.~M.~Figueroa-O'Farrill, B.~Spence and S.~Stanciu,
{\sl Planes, branes and automorphisms.  II: Branes in motion},
JHEP {\bf 9807}, 005 (1998) [hep-th/9805176].

\bibitem{Bandos:2001pu}
I.~A.~Bandos, J.~A.~de Azc{a}rraga, J.~M.~Izquierdo and J.~Lukierski,
{\sl BPS states in M-theory and twistorial constituents},
Phys.\ Rev.\ Lett.\ {\bf 86}, 4451 (2001) [hep-th/0101113].

\bibitem{Gauntlett:2002fz}
J.~P.~Gauntlett and S.~Pakis,
{\sl The geometry of $D = 11$ Killing spinors},
JHEP {\bf 0304}, 039 (2003) [hep-th/0212008].

\bibitem{Gauntlett:2003wb}
J.~P.~Gauntlett, J.~B.~Gutowski and S.~Pakis,
{\sl The geometry of D = 11 null Killing spinors},
JHEP {\bf 0312} (2003) 049.
[hep-th/0311112].

\bibitem{Gauntlett:2004zh}
J.~P.~Gauntlett, D.~Martelli, J.~Sparks and D.~Waldram,
{\sl Supersymmetric AdS(5) solutions of M-theory},
arXiv:hep-th/0402153.

\bibitem{Gauntlett:2002sc}
J.~P.~Gauntlett, D.~Martelli, S.~Pakis and D.~Waldram,
{\sl G-structures and wrapped NS5-branes},
Commun.\ Math.\ Phys.\  {\bf 247} (2004) 421.
[arXiv:hep-th/0205050].

\bibitem{Martelli:2003ki}
D.~Martelli and J.~Sparks,
{\sl G-structures, fluxes and calibrations in M-theory},
Phys.\ Rev.\ D {\bf 68}, 085014 (2003)
[arXiv:hep-th/0306225].

\bibitem{DLP1}
M. J. Duff, H. Lu and C. N. Pope,
{\sl Supersymmetry without supersymmetry},
Phys.\ Lett.\ B {\bf 409}, 136 (1997) [hep-th/9704186].

\bibitem{DLP2}
M. J. Duff, H. Lu and C. N. Pope,
{\sl $AdS_{5} \times S^{5}$ untwisted},
Nucl.\ Phys.\ B {\bf 532}, 181 (1998) [hep-th/9803061].

\bibitem{Supergravity81}
M. J. Duff,
{\sl Ultraviolet divergences in extended supergravity},
in Proceedings of the 1981 Trieste Conference ``Supergravity 81'', 
eds. Ferrara and Taylor, C. U. P. 1982.

\bibitem{Gunaydin1}
M. G{u}naydin,
{\sl Singleton and doubleton supermultiplets of space-time 
supergroups 
and infinite spin superalgebras}, in Proceedings of the 1989 Trieste 
Conference ``Supermembranes and Physics in $2+1$ Dimensions'', 
eds Duff, Pope and Sezgin, World Scientific 1990.

\bibitem{Sezgin}
E. Sezgin,
{\sl The spectrum of the eleven dimensional supergravity compactified
on the round seven sphere}, Trieste preprint, 1983, in Supergravity in
Diverse Dimensions, vol. 2, 1367, (eds A. Salam and E. Sezgin World 
Scientific, 1989);
Fortschr. Phys. {\bf 34} (1986) 217.

\bibitem{Duff:1999rk}
M.~J.~Duff,
{\sl TASI lectures on branes, black holes and anti-de Sitter space}, 
hep-th/9912164.

\bibitem{Aharony:1999ti}
O.~Aharony, S.~S.~Gubser, J.~M.~Maldacena, H.~Ooguri and Y.~Oz,
{\sl Large N field theories, string theory and gravity,'' Phys.\ Rept.\ 
{\bf 323}, 183 (2000)},
 [arXiv:hep-th/9905111].  

\bibitem{Maldacenatasi}
J.  Maldacena,
{\sl TASI lectures on AdS/CFT},
hep-th/0309246.

\bibitem{vanNwarner}
P. van Nieuwenhuizen and N. Warner,
{\sl Integrability conditions for Killing spinors},
Commun. Math. Phys. {\bf 93} (1984) 277.

\bibitem{Duffnilssonpopesuper}
M. J. Duff, B. E. W. Nilsson and C. N. Pope,
{\sl Spontaneous supersymmetry breaking by the squashed seven-sphere},
Phys.  Rev.  Lett.  {\bf 50} (1983) 2043.

\bibitem{Awadaduffpope}
M. A. Awada, M. J. Duff and C. N. Pope,
{\sl $N=8$ supergravity breaks down to $N=1$},
Phys.  Rev.  Lett.  {\bf 50} (1983) 294.

\bibitem{Castellani1}
L. Castellani, R.  D'Auria, P.  Fre and P.  van Nieuwenhuizen,
{\sl Holonomy groups, sesquidual torsion fields and $SU(8)$ in $d=11$
supergravity}, J. Math. Phys. {\bf 25} (1984) 3209. 

\bibitem{Lu:2003ze}
H.~Lu, C.~N.~Pope, K.~S.~Stelle and P.~K.~Townsend,
{\sl Supersymmetric deformations of $G_2$ manifolds from higher-order
corrections to string and M-theory},
hep-th/0312002.

\bibitem{Blencowe:1988sk}
M.~P.~Blencowe and M.~J.~Duff,
{\sl Supermembranes and the signature of space-time},
Nucl.\ Phys.\ B {\bf 310} (1988) 387.

\bibitem{Hull:1998ym}
C.~M.~Hull,
{\sl Duality and the signature of space-time}, JHEP {\bf 9811}(1998) 
017, hep-th/9807127.

\bibitem{Duffstelle}
M. J. Duff and K. Stelle,
{\sl Multimembrane solutions of $d = 11$ supergravity},
Phys.  Lett.  {\bf B253} (1991) 113.

\bibitem{U1}
M.~J.~Duff and J.~X.~Lu,
{\sl Duality Rotations In Membrane Theory},
Nucl.\ Phys.\ B {\bf 347}, 394 (1990).

\bibitem{U2}
C.~M.~Hull and P.~K.~Townsend,
{\sl Unity of superstring dualities},
Nucl.\ Phys.\ B {\bf 438}, 109 (1995)
[arXiv:hep-th/9410167].

\bibitem{West2}
P.~West,
{\sl The IIA, IIB and eleven dimensional theories and their common E(11) origin},
hep-th/0402140.

\bibitem{Batrachenko}
A.~Batrachenko and W.~Y.~Wen,
{\sl Generalized holonomy of supergravities with 8 real supercharges},
hep-th/0402141.



\end{thebibliography}
\end{document}